\documentclass[11pt,letter]{article}  

\def\statusstring{Submitted to IEEE Transactions on Information Theory,
                  September 2006}

\usepackage{amsthm} 
\usepackage{wrapfig}
\usepackage{amssymb}
\usepackage{latexsym}
\usepackage{epsfig}
\usepackage{graphicx}
\usepackage{amsmath}
\usepackage{epsfig}
\usepackage{multicol}
\usepackage{psfrag}
\usepackage{float}
\usepackage{cite}
\usepackage{subfigure}

\def\mod{\operatorname{mod}\,}

\newcommand{\setV}{\set{V}}

\newcommand{\wpsAWGNCmin}{w_{\mathrm{p}}^{\mathrm{AWGNC,min}}}
\newcommand{\wpsBSCmin}{w_{\mathrm{p}}^{\mathrm{BSC,min}}}
\newcommand{\wpsBECmin}{w_{\mathrm{p}}^{\mathrm{BEC,min}}}

\newcommand{\wpsminh}[1]{\wpsmin(\matr{#1})}
\newcommand{\wpsAWGNCminh}[1]{\wpsAWGNCmin(\matr{#1})}
\newcommand{\wpsBSCminh}[1]{\wpsBSCmin(\matr{#1})}
\newcommand{\wpsBECminh}[1]{\wpsBECmin(\matr{#1})}

\newcommand{\vomegabar}{\overline{\boldsymbol{\omega}}}
\newcommand{\dmin}{d_{\mathrm{min}}}
\newcommand{\dfree}{d_{\mathrm{free}}}
\newcommand{\code}[1]{\mathsf{#1}}
\newcommand{\matr}[1]{\mathbf{#1}}
\newcommand{\vect}[1]{\mathbf{#1}}
\newcommand{\graph}[1]{\mathcal{#1}}
\newcommand{\set}[1]{\mathcal{#1}}
\newcommand{\GF}[1]{\mathbb{F}_{#1}}
\newcommand{\defeq}{\triangleq}

\newcommand{\tr}{\mathsf{T}}
\newcommand{\ringFXr}{\GF{2}[X] / \langle X^r - 1 \rangle}
\newcommand{\ringFXtwor}{\GF{2}[X] / \langle X^{2r} - 1 \rangle}
\newcommand{\dint}[1]{\,\operatorname{d}{#1}}

\newtheorem{theorem}{Theorem}[section]
\newtheorem{lemma}[theorem]{Lemma}

\theoremstyle{definition}
\newtheorem{definition}[theorem]{Definition}

\newtheorem{remark}[theorem]{Remark}
\numberwithin{equation}{section}

\newtheorem{PreExample}[theorem]{{\textbf{Example}}}
  \newenvironment{Example}%
    {\begin{PreExample}}{\hfill$\square$\end{PreExample}}

\newcommand{\PGq}{\operatorname{PG}(2,q)}

\newcommand{\supp}{\operatorname{supp}}

\newcommand{\vlambda}{\boldsymbol{\lambda}}
\newcommand{\vomega}{\boldsymbol{\omega}}

\newcommand{\vc}{\vect{c}}

\newcommand{\vubar}{\overline{\vect{u}}}

\newcommand{\Rp}{\mathbb{R}_{+}}
\newcommand{\Rpp}{\mathbb{R}_{++}}

\newcommand{\wfr}{w_{\mathrm{frac}}}
\newcommand{\wfrmin}{w^{\mathrm{min}}_{\mathrm{frac}}}
\newcommand{\wfrminh}[1]{w^{\mathrm{min}}_{\mathrm{frac}}(\matr{#1})}

\newcommand{\wmaxfr}{w_{\mathrm{max-frac}}}
\newcommand{\wmaxfrminh}[1]{w^{\mathrm{min}}_{\mathrm{max-frac}}(\matr{#1})}
\newcommand{\wmaxfrmin}{w^{\mathrm{min}}_{\mathrm{max-frac}}}
\newcommand{\wmaxfrminlr}[1]{w^{\mathrm{min}}_{\mathrm{max-frac}}\left( #1 \right)}

\newcommand{\edefinition}{\hfill$\square$}

\newcommand{\convhull}{\operatorname{ConvexHull}}

\newcommand{\fp}[1]{\set{#1}}
\newcommand{\fph}[2]{\set{#1}(\matr{#2})}
\newcommand{\fc}[1]{\set{#1}}
\newcommand{\fch}[2]{\set{#1}(\matr{#2})}

\newcommand{\wps}{w_{\mathrm{p}}}
\newcommand{\wpsAWGNC}{w_{\mathrm{p}}^{\mathrm{AWGNC}}}

\newcommand{\wpsBSC}{w_{\mathrm{p}}^{\mathrm{BSC}}}
\newcommand{\wpsBEC}{w_{\mathrm{p}}^{\mathrm{BEC}}}
\newcommand{\wpsmin}{w_{\mathrm{p}}^{\mathrm{min}}}
\newcommand{\infnorm}[1]{\lVert #1 \rVert_{\infty}}
\newcommand{\infnormbig}[1]{\big\lVert #1 \big\rVert_{\infty}}
\newcommand{\onenorm}[1]{\lVert #1 \rVert_1}

\newcommand{\twonorm}[1]{\lVert #1 \rVert_2}
\newcommand{\twonormbig}[1]{\big\lVert #1 \big\rVert_2}
\newcommand{\wH}{w_{\mathrm{H}}}
\newcommand{\wHmin}{w_{\mathrm{H}}^{\mathrm{min}}}

\newcommand{\codeCconv}{\code{C}_{\mathrm{conv}}}
\newcommand{\codeCQC}[1]{\code{C}_{\mathrm{QC}}^{(r)}}
\newcommand{\matrGconv}{\matr{G}_{\mathrm{conv}}}
\newcommand{\matrGbarconv}{\overline{\matr{G}}_{\mathrm{conv}}}
\newcommand{\matrHconv}{\matr{H}_{\mathrm{conv}}}
\newcommand{\matrHbarconv}{\overline{\matr{H}}_{\mathrm{conv}}}
\newcommand{\matrHQC}[1]{\matr{H}_{\mathrm{QC}}^{(#1)}}
\newcommand{\matrHbarQC}[1]{\overline{\matr{H}}_{\mathrm{QC}}^{(#1)}}
\newcommand{\setEconv}{\set{E}_{\mathrm{conv}}}

\newcommand{\setEQC}[1]{\set{E}_{\mathrm{QC}}^{(#1)}}

\newcommand{\vectbar}[1]{\overline{\vect{#1}}}

\newcommand{\codebar}[1]{\overline{\code{#1}}}
\newcommand{\matrOmega}{\boldsymbol{\Omega}}

\newcommand{\R}{\mathbb{R}}

\newenvironment{pr}{{\sc Proof:}}{\mbox{}\hfill$\Box$\par}

\newcommand{\matrKconv}{\matr{K}_{\mathrm{conv}}}
\newcommand{\matrKQC}[1]{\matr{K}_{\mathrm{QC}}^{(#1)}}
\newcommand{\codeCQCtwor}{\code{C}_{\mathrm{QC}}^{(2r)}}
\newcommand{\codeCQCfourr}{\code{C}_{\mathrm{QC}}^{(4r)}}

\newcommand{\me}{m_{\mathrm{e}}}
\newcommand{\ms}{m_{\mathrm{s}}}
\newcommand{\dc}{d^{\mathrm{c}}}
\newcommand{\dr}{d^{\mathrm{r}}}

\newcommand{\Eb}{E_{\mathrm{b}}}
\newcommand{\Es}{E_{\mathrm{s}}}

\begin{document}



\title{Pseudo-Codeword Performance Analysis \\
       for LDPC Convolutional Codes%
  \footnote{The first author was partially supported by NSF Grant
      CCR-0205310. 
      The second and fourth authors were partially supported by NSF Grants
      CCR-0205310 and CCF-0514801, and by NASA Grant NNG05GH73G.
      While at MIT, the third author was partially supported by NSF
      Grants CCF-0514801 and CCF-0515109 and 
      by HP through the MIT/HP Alliance.
      Some of the material in this paper was presented at the 2006 IEEE 
      International Symposium on Information Theory, 
      see~\cite{Smarandache:Pusane:Vontobel:Costello:06:1}.}  }

\author{Roxana Smarandache,%
  \footnote{Department of Mathematics and Statistics, San Diego State
            University, San Diego, CA 92182, USA. The work for this paper was
            partially done while on leave at the Department of
            Mathematics, University of Notre Dame, Notre Dame, IN 46556, USA.
            Email: \texttt{rsmarand@sciences.sdsu.edu}.
            R.~Smarandache is the corresponding author.}
         \ Ali E. Pusane,%
    \footnote{Department of EE, University of Notre Dame, Notre Dame,
            IN 46556, USA. Email: \texttt{apusane@nd.edu}.}
         \ Pascal O.~Vontobel,%
    \footnote{Hewlett-Packard Laboratories,
              1501 Page Mill Road, Palo Alto, CA 94304, USA. 
              E-Mail: \texttt{pascal.vontobel@ieee.org}. The work for this
              paper was partially done while at the Dept.~of EECS,
              Massachusetts Institute of Technology, Cambridge, MA, USA.}
         \ and Daniel J.~Costello, Jr.%
    \footnote{Department of EE, University of Notre Dame, Notre Dame, 
              IN 46556, USA.  Email:
              \texttt{costello2@nd.edu}.}
    }

\date{}

\maketitle

\vspace{-5.75cm}
{
 \begin{flushright}
   \texttt{\statusstring}
 \end{flushright}
}
\vspace{+5cm}


\begin{abstract}
  Message-passing iterative decoders for low-density parity-check (LDPC) block
  codes are known to be subject to decoding failures due to so-called
  pseudo-codewords. These failures can cause the large signal-to-noise ratio
  performance of message-passing iterative decoding to be worse than that
  predicted by the maximum-likelihood decoding union bound.
  
  In this paper we address the pseudo-codeword problem from the
  convolutional-code perspective. In particular, we compare the performance of
  LDPC convolutional codes with that of their ``wrapped'' quasi-cyclic block
  versions and we show that the minimum pseudo-weight of an LDPC convolutional
  code is at least as large as the minimum pseudo-weight of an underlying
  quasi-cyclic code. This result, which parallels a well-known relationship
  between the minimum Hamming weight of convolutional codes and the minimum
  Hamming weight of their quasi-cyclic counterparts, is due to the fact that
  every pseudo-codeword in the convolutional code induces a pseudo-codeword in
  the block code with pseudo-weight no larger than that of the convolutional
  code's pseudo-codeword. This difference in the weight spectra leads to
  improved performance at low-to-moderate signal-to-noise ratios for the
  convolutional code, a conclusion supported by simulation results.
\end{abstract}

\noindent\textbf{Index terms} --- 
Convolutional codes, 
quasi-cyclic codes, 
low-density parity-check (LDPC) codes, 
linear programming decoding, 
message-passing iterative decoding,
pseudo-codewords, 
pseudo-weights.
 

\newpage

\section{Introduction}
\label{sec:introduction}

\begin{figure}
  \begin{center}
    \epsfig{file=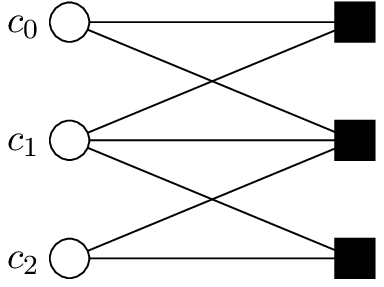, height=3.5cm}
    \hspace{2cm}
    \epsfig{file=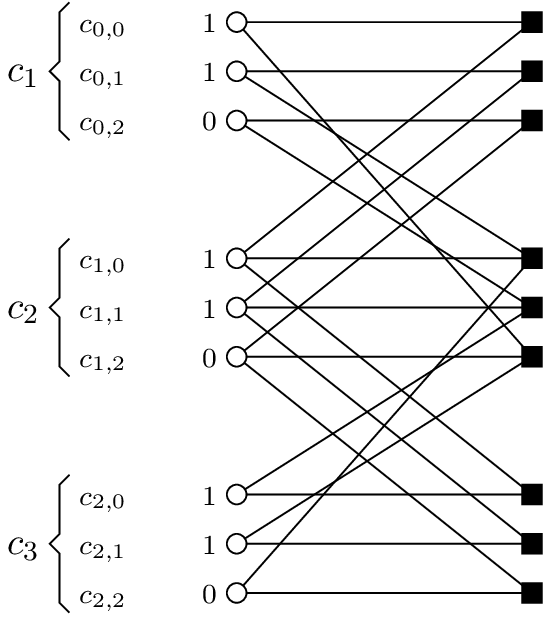, height=3.5cm}
  \end{center}
  \caption{Left: the Tanner graph associated with the parity-check matrix
    $\matr{H}$ that was defined in \eqref{eq:pcm:trivial:code:1}.  Right:
    $(1{:}1{:}0, \ 1{:}1{:}0, \ 1{:}1{:}0)$ is a valid codeword in a cubic
    cover.}
  \label{fig:smallgraph:1} 
\end{figure}

Although low-density parity-check block codes (LDPC-BCs) have very good
performance under message-passing iterative (MPI) decoding, they are known to
be subject to decoding failures due to so-called pseudo-codewords. These are
real-valued vectors that can be loosely described as error patterns that cause
non-convergence in iterative decoding due to the fact that the algorithm works
locally and can give priority to a vector that fulfills the equations of a
graph cover rather than the graph itself. 
\begin{Example} \label{example:introduction} Consider the trivial
length-$3$ and dimension-$0$ code $\code{C} = \{ (0, 0, 0) \}$ 
defined by the $3 \times 3$ parity-check matrix
\begin{align}
  \matr{H}
    &\defeq
       \begin{bmatrix}
         1 & 1 & 0 \\
         1 & 1 & 1 \\
         0 & 1 & 1
       \end{bmatrix}
         \label{eq:pcm:trivial:code:1}
\end{align}
and whose Tanner graph is shown in Fig.~\ref{fig:smallgraph:1}~(left). A
possible cubic cover of this Tanner graph is also depicted in
Fig.~\ref{fig:smallgraph:1} (right). Because $(1{:}1{:}0, \ 1{:}1{:}0, \ 1{:}1{:}0)$
is a valid configuration in this cubic cover, the vector $\vomega = \bigl(
\frac{2}{3}, \frac{2}{3}, \frac{2}{3} \bigr)$ is a pseudo-codeword of the code
$\code{C}$ defined by $\matr{H}$ (see~\cite{Koetter:Vontobel:03:1,
  Vontobel:Koetter:05:1:subm}).
\end{Example}

It has been shown~\cite{Koetter:Vontobel:03:1, Wiberg:96,
  Forney:Koetter:Kschischang:Reznik:01:1} that the performance of MPI decoding
schemes for LDPC-BCs is largely dominated \emph{not} by minimum Hamming weight
considerations but by minimum \emph{pseudo-weight} considerations, where the
minimum pseudo-weight in the case of an additive white Gaussian noise channel
(AWGNC) is defined as $\wpsminh{H} \defeq \min_{\vomega \in \mathcal P\atop
  \vomega \neq \vect{0} } \frac{\onenorm{\vomega}^2} {\twonorm{\vomega}^2},$
where $\onenorm{\,\cdot\,}$ and $\twonorm{\,\cdot\,}$ are, respectively, the
$1$-norm and $2$-norm, and $\mathcal P$ is the set of all pseudo-codewords for
the code $\code{C}$ defined by $\matr{H}$. The minimum pseudo-weight is a
measure of the effect that decoding failures have on the performance of the
code.\footnote{Since $\wpsminh{H} \leq \wHmin(\code{C})$, the minimum Hamming
  weight may still be a key factor in performance analysis; however, the
  larger the gap between $\wpsminh{H}$ and $\wHmin(\code{C})$, the greater
  role the minimum pseudo-weight plays. In any case, the minimum Hamming
  weight is important for quantifying the impact of undetectable errors.} As a
consequence, the large signal-to-noise ratio (SNR) performance of MPI decoding
can be worse than that predicted by the maximum-likelihood decoding union
bound, which constitutes a major problem when trying to determine performance
guarantees. Addressing this problem from the convolutional-code perspective,
i.e., studying the pseudo-codeword problem described above for LDPC
convolutional codes (LDPC-CCs), constitutes the major topic of this paper.

\begin{figure}
  \begin{center}
    \epsfig{file=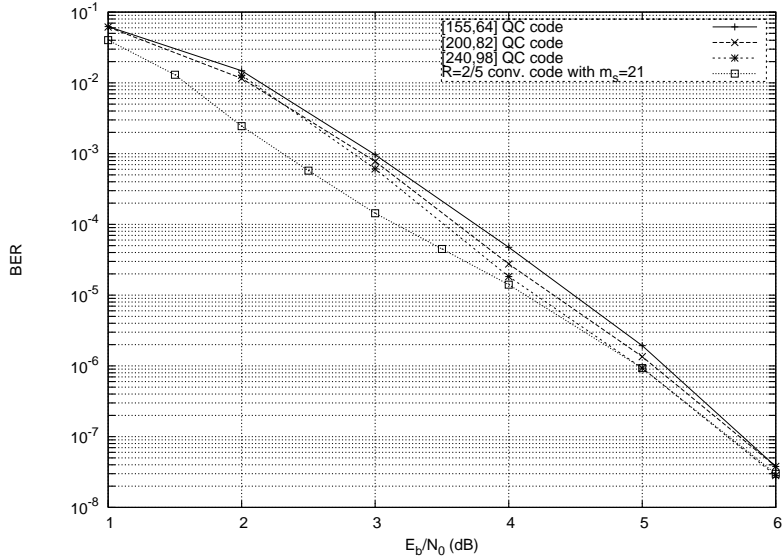, width=0.65\columnwidth}
  \end{center}
  \caption{The MPI decoding performance of a rate $R=2/5$ $(3,5)$-regular LDPC
    convolutional code and three associated $(3,5)$-regular QC-LDPC block
    codes. (Maximum of $50$ iterations.)}
  \label{fig:compare:1}
\end{figure}

We investigate a class of time-invariant LDPC convolutional codes derived by
``unwrapping'' certain classes of quasi-cyclic (QC) LDPC block codes that are
known to have good performance
\cite{Tanner:Sridhara:Sridharan:Fuja:Costello:04:1,
  Sridharan:Costello:Sridhara:Fuja:Tanner:02:1, Smarandache:Vontobel:04:1}.
Unwrapping a QC block code to obtain a time-invariant convolutional code
represents a major link between QC block codes and convolutional codes. This
link was first introduced in a paper by Tanner~\cite{Tanner:87:1}, where it
was shown that the free distance of the unwrapped convolutional code, if
non-trivial, cannot be smaller than the minimum distance of the underlying QC
code. This idea was later extended in~\cite{Levy:Costello:93:1,
  Esmaeili:Gulliver:Secord:Mahmoud:98:1}.  More recently, a construction for
LDPC convolutional codes based on QC-LDPC block codes was introduced by Tanner
et al.~\cite{Sridharan:Costello:Sridhara:Fuja:Tanner:02:1,
  Tanner:Sridhara:Sridharan:Fuja:Costello:04:1}, and a sliding-window MPI
decoder was described. In that paper it was noted that the (non-trivial)
convolutional versions of these codes significantly outperformed their block
code counterparts in the waterfall region of the bit error rate (BER) curve,
even though the graphical representations of MPI decoders were essentially
equivalent.

Throughout this paper we mainly take the approach
of~\cite{Koetter:Vontobel:03:1, Vontobel:Koetter:05:1:subm,
Vontobel:Koetter:04:2}, which connects the presence of
pseudo-codewords in MPI decoding and linear programming (LP)
decoding. LP decoding was introduced by Feldman, Wainwright, and
Karger~\cite{Feldman:03:1, Feldman:Wainwright:Karger:05:1} and can be
seen as a relaxation of the maximum-likelihood decoding problem. More
precisely, maximum-likelihood decoding can be formulated as the
solution to an optimization problem, where a linear cost function is
minimized over a certain polytope, namely the polytope that is spanned
by the set of all codewords. For general codes, there is no efficient
description of this polytope and so Feldman et al.~suggested 
replacing it with an (efficiently describable) relaxed polytope, which in
the following will be called the ``fundamental polytope''. In other
words, the decoding result of the LP decoder is the point in the
fundamental polytope that minimizes the above-mentioned linear cost
function.

In order to analyze the behavior of unwrapped LDPC convolutional codes under
LP decoding, we need to examine the fundamental
polytope~\cite{Koetter:Vontobel:03:1, Vontobel:Koetter:05:1:subm} of the
underlying QC-LDPC block codes. (Because of symmetries, it is actually
sufficient to study the structure of the fundamental polytope around the zero
codeword, i.e., it is sufficient to study the so-called fundamental cone.) Our
goal is to formulate analytical results (or at least efficient procedures)
that will allow us to bound the minimum pseudo-weight of the pseudo-codewords
of the block and convolutional codes.

The paper aims at addressing this question and related issues. In the
following sections, we will study the connections that exist between
pseudo-codewords in QC codes and pseudo-codewords in the
associated convolutional codes and show that this connection mimics the
connection between the codewords in QC codes and the associated
convolutional codes. 

\subsection{Motivational Example}
As a motivational example we simulated a rate $R = 2/5$ $(3,5)$-regular
LDPC-CC with syndrome former memory $m_\text{s} = 21$, together with three
wrapped block code versions: a $[155,64]$ $(3,5)$-regular QC-LDPC block code,
a $[200,82]$ code, and a $[240,98]$ code, with parity-check matrices of
increasing circulant sizes $r = 31$, $r = 40$, and $r = 48$, respectively,
while keeping the same structure within each $r \times r$
circulant~\cite{Smarandache:Pusane:Vontobel:Costello:06:1} (see
Section~\ref{sec:link:qc:and:conv:codes:1}).  (Note that increasing the
circulant size of the QC code increases its complexity, i.e., its block
length. Also note that each of the three block codes has rate slightly greater
than $2/5$.) A sliding-window MPI decoder as
in~\cite{Tanner:Sridhara:Sridharan:Fuja:Costello:04:1} was used to decode the
convolutional code. Conventional LDPC-BC MPI decoders were employed to decode
the QC-LDPC block codes. All decoders were allowed a maximum of $50$
iterations.  The resulting BER performance of these codes on a binary-input
AWGN channel is shown in Fig.~\ref{fig:compare:1}.

\begin{figure}
  \begin{center}
    \epsfig{file=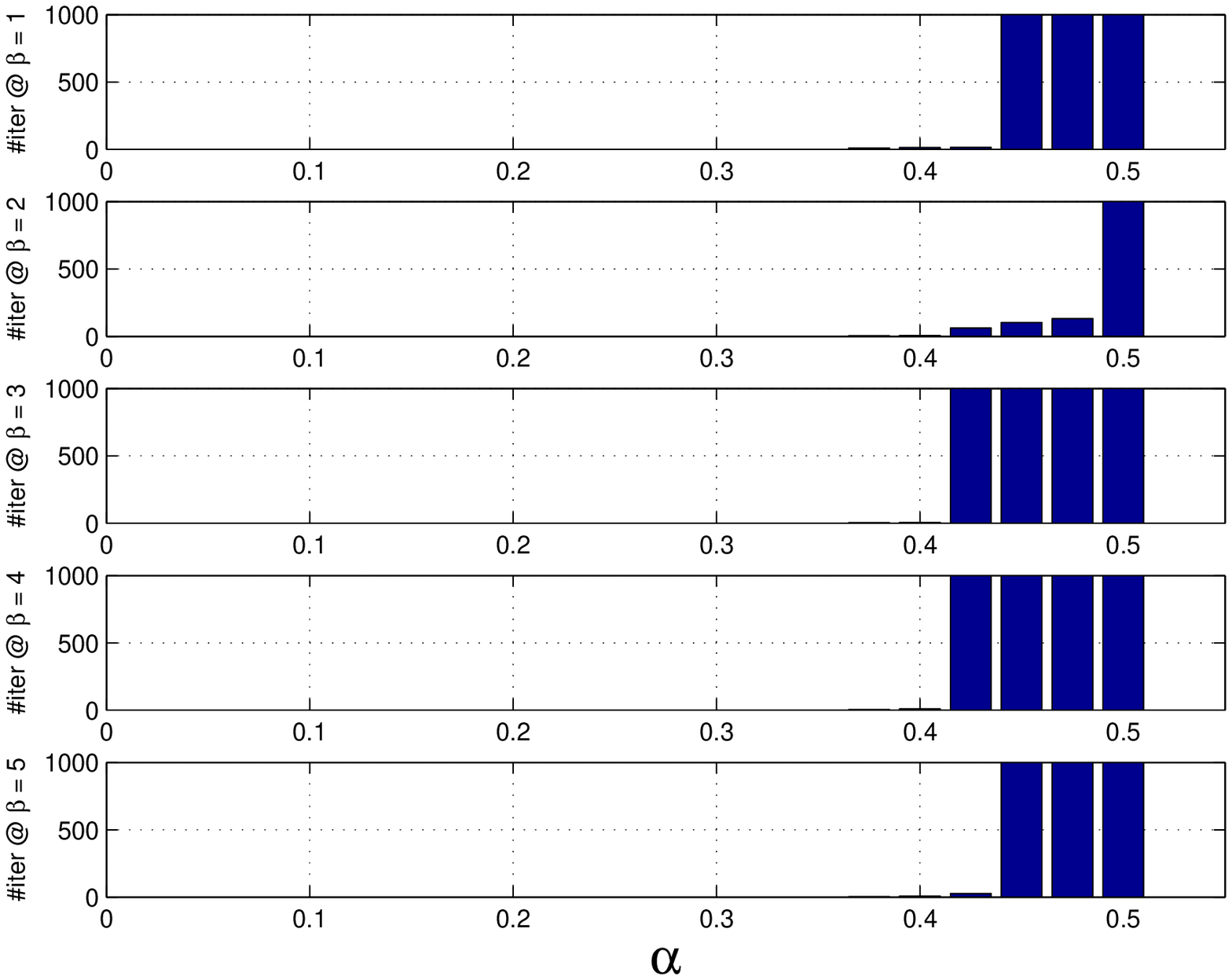, width=7cm}
    \quad
    \epsfig{file=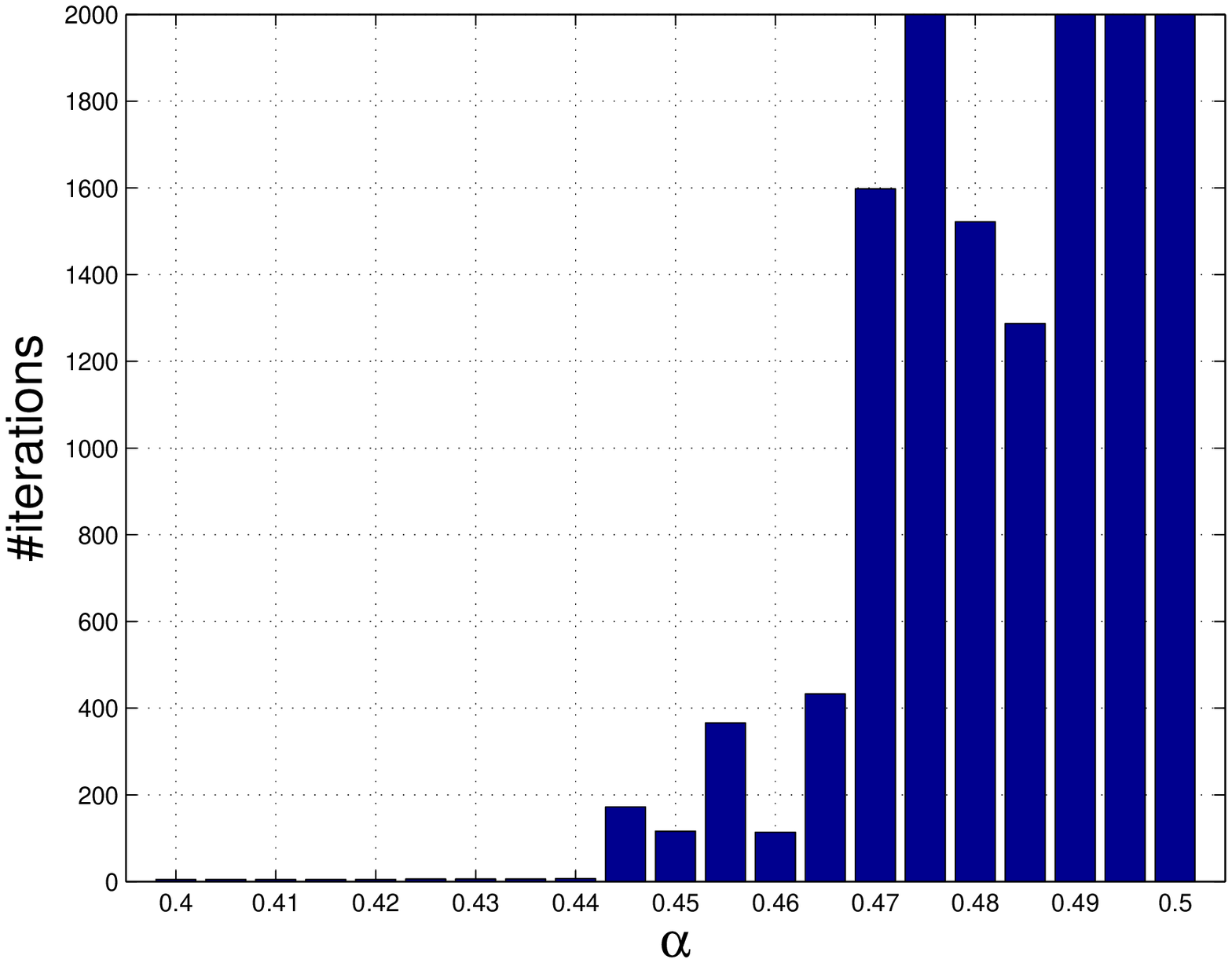, width=7cm}
  \end{center}
  \caption{Left: Number of iterations needed for a sum-product-algorithm-type
    MPI decoder to decide for the all-zero codeword. ($1000$ corresponds to no
    convergence.) Here the $\beta$ values $1.00$, $2.00$, $3.00$, $4.00$,
    $5.00$ correspond to signal-to-noise ratios $\Eb/N_0$ of, respectively,
    $-2.04 \ \mathrm{dB}$, $0.96 \ \mathrm{dB}$, $2.73 \ \mathrm{dB}$, $3.97 \ 
    \mathrm{dB}$, $4.94 \ \mathrm{dB}$. Right: Number of iterations needed for
    a min-sum-algorithm-type MPI decoder to decide for the all-zero codeword.
    ($2000$ corresponds to no convergence.) We did not observe convergence to
    the all-zero codeword for $\alpha > 0.50$ either for the
    sum-product-algorithm-type MPI decoder or for the min-sum-algorithm-type
    MPI decoder.}
  \label{fig:alphabetaplot:1} 
\end{figure}

We note that, particularly in the low-to-moderate SNR region, where the
complete pseudo-weight spectrum plays an important role, the unwrapped LDPC-CC
performs between $0.5 \ \mathrm{dB}$ and $1.0 \ \mathrm{dB}$ better than the
associated QC-LDPC block codes. Also, as the circulant size increases, the
performance of the block codes approaches that of the convolutional code.
These performance curves suggest that the pseudo-codewords in the block code
that result in decoding failures may not cause such failures in the
convolutional code,
which suggests that LDPC-CCs may have better iterative decoding thresholds
than comparable LDPC-BCs (see
also~\cite{Sridharan:Lentmaier:Zigangirov:Costello:04:1}
and~\cite{Huebner:Lentmaier:Zigangirov:Costello:05:1}).

In order to underline the influence of pseudo-codewords under MPI decoding, we
consider the following experiment. Let $\vomega$ be a {\em minimal
  pseudo-codeword}~\cite{Vontobel:Koetter:05:1:subm} for the above-mentioned
$R = 2/5$ $(3,5)$-regular LDPC-CC, i.e., a pseudo-codeword that corresponds to
an {\em edge of the fundamental cone} of that LDPC-CC. Moreover, we define the
log-likelihood ratio vector $\vlambda(\alpha, \beta)$ to be\footnote{Note that
  the numerator in the ratio $\onenorm{\vomega} / \twonorm{\vomega}^2$ is not
  squared, therefore this ratio does not correspond to the AWGNC pseudo-weight
  of $\vomega$ (see Sec.~\ref{sec:pseudo:weight:comparison:1}), although it is
  closely related to that value.}
\begin{align*}
  \vlambda(\alpha, \beta) 
    &\defeq
       \beta \cdot
       \left(
         \vect{1}
         - 
         2 
         \alpha 
         \frac{\onenorm{\vomega}}
              {\twonorm{\vomega}^2}
         \vomega
       \right).
\end{align*}
We then run the MPI decoder that is initialized with $\vlambda =
\vlambda(\alpha, \beta)$ and count how many iterations it takes until the
decoder decides for the all-zero codeword as a function of $\alpha$ and
$\beta$.  The results are shown in Fig.~\ref{fig:alphabetaplot:1}.

The meaning of $\vlambda(\alpha, \beta)$ is the following. If $\alpha = 0$,
then $\vlambda$ corresponds to the log-likelihood ratio vector that the
receiver sees when the communication system operates at a signal-to-noise
ratio $\Eb / N_0 = \beta / (4R)$ and when the noise vector that is added by
the binary-input AWGN channel happens to be the all-zero vector (see,
e.g., the discussion in~\cite[Sec.~3]{Vontobel:Koetter:05:1:subm}). For
non-zero $\alpha$ the expression for $\vlambda(\alpha,\beta)$ has been set up
such that the LP decoder has a decision boundary at $\alpha = 0.5$: for
$\alpha < 0.5$ the all-zero codeword wins against the pseudo-codeword
$\vomega$ whereas for $\alpha > 0.5$ the all-zero codeword loses against the
pseudo-codeword $\vomega$ under LP decoding.

The simulations in Fig.~\ref{fig:alphabetaplot:1} were obtained using a search
algorithm that looked for a low-pseudo-weight minimal pseudo-codeword
$\vomega$ in the fundamental cone of the above-mentioned LDPC-CC. The
$\vomega$ that we found has AWGNC pseudo-weight $18.1297$, which happens to be
smaller than the free distance.  Secondly, we ran the MPI decoder for various
choices of $\alpha$ and $\beta$: Fig.~\ref{fig:alphabetaplot:1}~Left shows the
number of iterations needed using a sum-product-algorithm-type MPI decoder
whereas Fig.~\ref{fig:alphabetaplot:1}~Right shows the number of iterations
needed using a min-sum-algorithm-type MPI decoder. (Note that the decisions
reached by the latter are independent of the choice of $\beta$, $\beta > 0$.)
Because of the more oscillatory behavior of the min-sum-algorithm-type MPI
decoder close to decision boundaries of the LP decoder, observed empirically,
it is advantageous to run that decoder for many iterations in our scenario,
whereas in the case of the sum-product-algorithm-type MPI decoder it hardly
pays  to go beyond $150$ iterations.

\subsection{Paper Goals and Structure} 
In this paper, we provide a possible explanation for the performance
difference observed in the motivational example above. Based on the
results of~\cite{Koetter:Vontobel:03:1, Vontobel:Koetter:05:1:subm}
that relate code performance to the existence of pseudo-codewords, we
examine the pseudo-codeword weight spectra of QC-LDPC block codes and
their associated convolutional codes. We will show that for a
non-trivial LDPC-CC derived by unwrapping a non-trivial QC-LDPC block
code,\footnote{Non-trivial means here that the set of pseudocodewords
contain non-zero pseudo-codewords.} the minimum pseudo-weight of the
convolutional code is at least as large as the minimum pseudo-weight
of the underlying QC
code\cite{Smarandache:Pusane:Vontobel:Costello:06:1}, i.e.,
\begin{align*}
  \wpsmin\left( \matrHQC{r} \right)
    &\leq \wpsmin\left( \matrHconv \right).
\end{align*}
  
This result, which parallels the well-known relationship between the
free Hamming distance of non-trivial convolutional codes and the
minimum Hamming distance of their non-trivial quasi-cyclic
counterparts~\cite{Tanner:87:1},\footnote{Non-trivial means here that
the set of codewords contain non-zero codewords.} is based on the fact
that every pseudo-codeword in the convolutional code induces a
pseudo-codeword in the block code with pseudo-weight no larger than
that of the convolutional code's pseudo-codeword. This difference in
the weight spectra leads to improved BER performance at
low-to-moderate SNRs for the convolutional code, a conclusion
supported by the simulation results presented in
Fig.~\ref{fig:compare:1}.


The paper is structured as follows. In Sec.~\ref{sec:ps:qc:and:conv:codes:1}
we develop the background necessary to describe the connection between
pseudo-codewords in unwrapped LDPC convolutional codes and those in the
associated QC-LDPC codes. Thus, in Sec.~\ref{sec:link:qc:and:conv:codes:1} we
briefly discuss the connection between convolutional codes and their
associated QC codes, especially how codewords in the former can be used to
construct codewords in the latter, and in
Sec.~\ref{sec:fundam:cone:par:check:matrix:1} we define the fundamental
polytope/cone of a matrix and show how we can describe the fundamental cone of
a polynomial parity-check matrix through polynomial inequalities. We end the
section by showing how pseudo-codewords in unwrapped LDPC convolutional codes
yield pseudo-codewords in the associated QC-LDPC codes.  In
Sec.~\ref{sec:pseudo:weight:comparison:1}, we compare the pseudo-weights of
unwrapped convolutional and their associated QC block codes.
Sec.~\ref{sec:pseudo:weight:definitions:1} introduces various channel
pseudo-weights and Sec.~\ref{sec:pseudo:weight:inequality:1} presents the main
result, namely that the minimum AWGN pseudo-weight, the minimum binary erasure
channel (BEC) pseudo-weight, the minimum binary symmetric channel (BSC)
pseudo-weight, and the minimum max-fractional weight of a convolutional code
are at least as large as the corresponding minimum pseudo-weights of a wrapped
QC block code.
Sec.~\ref{sec:analysis:problematic:pseudo:codewords:conv:codes:1} discusses a
method to analyze problematic pseudo-codewords, i.e., pseudo-codewords with
small pseudo-weight. The method addresses the convolutional code case. It
introduces two sequences of ``truncated'' pseudo-weights and, respectively,
``bounded pseudo-codeword'' pseudo-weights, that play an important role in
identifying the minimum pseudo-weight of the convolutional code,
similar to the role that column distances and row distances play in
identifying the free distance.
We end with some conclusions in Sec.~\ref{sec:conclusions:1}.

\subsection{Notation}
Throughout the paper we will use the standard way of associating a Tanner
graph with a parity-check matrix and vice-versa.  We will use the following
notation. We let $\GF{2}$, $\R$, $\Rp$, and $\Rpp$ be the Galois field of size
$2$, the field of real numbers, the set of non-negative real numbers, and the
set of positive real numbers, respectively.  If $a(X)$ is a polynomial over
some field and $r$ is some positive integer then $\big( a(X) \mod(X^r-1)
\big)$ denotes the polynomial $b(X)$ of degree smaller than $r$ such that
$b(X) = a(X) \ (\mathrm{mod}\ (X^r-1))$. We say that a polynomial
$\omega_{\ell}(D) = \sum_{i} \omega_{\ell, i} D^i$ with real coefficients is
non-negative, and we write $\omega_{\ell}(D) \geq 0$, if all its coefficients
$\omega_{\ell,i}$ satisfy $\omega_{\ell,i} \geq 0$.  Similarly, a polynomial
vector $\vomega(D) = \bigl( \omega_1(D), \omega_2(D), \ldots, \omega_L(D)
\bigr)$ is non-negative, and we write $\vomega(D) \geq \vect{0}$, if all its
polynomial components $\omega_{\ell}(D)$ satisfy $\omega_{\ell}(D) \geq 0$ for
all $\ell \in \{ 1, \ldots, L \}$.  Moreover, a polynomial matrix
$\matr{A}(D)$ is non-negative, and we write $\matr{A}(D) \geq \matr{0}$, if
all its entries are non-negative polynomials. Finally, for any positive
integer $r$ and for any $r' \in \{ 0, 1, \ldots, r-1 \}$,  $\matr{I}_{r'}$ will
represent the $r'$-times cyclically left-shifted identity matrix of size $r
\times r$.


\section{Pseudo-Codewords for QC and 
            Convolutional Codes}

\label{sec:ps:qc:and:conv:codes:1}


We begin this section by presenting an important link between QC block codes
and convolutional codes that was first introduced in a paper by
Tanner~\cite{Tanner:87:1} and later extended in~\cite{Levy:Costello:93:1,
  Esmaeili:Gulliver:Secord:Mahmoud:98:1}. Similar to the connection between
codewords in an unwrapped convolutional code and codewords in the underlying
QC code, we will show  
that pseudo-codewords in the convolutional code give pseudo-codewords when
projected onto an underlying QC code.


\subsection{A Link Between QC Block Codes and Convolutional Codes}

\label{sec:link:qc:and:conv:codes:1}


In this section we introduce the background needed for the later development
of the paper. Note that all codes will be binary linear codes.



Any length $n\defeq r \cdot L$ quasi-cyclic (QC) code $\code{C}_{\rm QC}
\defeq \codeCQC{r}$ with period $L$ can be represented by an $rJ\times rL$
{\em scalar} block parity-check matrix $\matrHQC{r}$ that consists of $J
\times L$ circulant matrices of size $r \times r$.  Using the isomorphism between
the ring of binary circulant matrices of size $r \times r$ and the ring of
polynomials $\ringFXr$ of degree less than $r$, we can associate with the
scalar parity-check matrix $\matrHQC{r}$ the \emph{polynomial} parity-check
matrix $\matrHQC{r}(X) \in \big( \ringFXr \big)^{J \times L}$, with polynomial
operations performed modulo $X^r-1$.  Due to the existence of this
isomorphism, we can identify two descriptions, scalar and polynomial, and use
either of the two depending on their usefulness.

\begin{Example}\label{cubic:matrices}
  The cubic cover $\code{C^{(3)}}$ of the trivial code in depicted in
  Fig.~\ref{fig:smallgraph:1} is a quasi-cyclic code\footnote{Not all covers
    are QC codes.} with  scalar parity-check matrix $\matrHQC{3}$ and 
  corresponding polynomial parity-check matrix $\matrHQC{3}(X)\in \big(
  \ringFXr \big)^{3 \times 3}$ ($r=3$) given by

{\small \begin{align} \matrHQC{3}=
      & \begin{bmatrix}
        1& 0 &0& 1 &0 &0&0 &0 &0 \\ 0& 1 &0& 0 &1 &0& 0 &0 &0\\
        0& 0 &1& 0 &0 &1&0 &0 &0 \\  0& 1 &0& 1 &0 &0& 0 &0 &1 \\
        0& 0 &1& 0 &1 &0&1 &0 &0 \\ 1& 0 &0& 0 &0 &1& 0 &1 &0 \\
        0& 0 &0& 1 &0 &0&1 &0 &0 \\ 0& 0 &0& 0 &1 &0& 0 &1 &0 \\
        0& 0 &0& 0 &0 &1&0 &0 &1
       \end{bmatrix}= 
      \begin{bmatrix}
          \matr{I}_{0}& \matr{I}_{0} & \matr{0}\\
          \matr{I}_{2}& \matr{I}_{0} & \matr{I}_{1}\\
          \matr{0}  & \matr{I}_{0} & \matr{I}_{0}
       \end{bmatrix},  \matrHQC{3}(X)= \begin{bmatrix}
          1& 1 & 0\\
          X^2& 1 & X\\
          0  & 1 & 1
       \end{bmatrix}.
         \label{eq:pcm:cobiccover:code:1}
\end{align}}
\end{Example}
By permuting the rows and columns of the scalar parity-check matrix
$\matrHQC{r}$ (i.e., by taking the first row in the first block of $r$ rows, the first row of
the second block of $r$ rows, etc., then the second row in the first block, the second row
in the second block, etc., and similarly for the columns), we obtain the
parity-check matrix $\matrHbarQC{r}$ of a code that is equivalent to
$\codeCQC{r}$. The scalar parity-check matrix
$\matrHbarQC{r}$ has the form
\begin{align*}
  \matrHbarQC{r}
    &\defeq
       \begin{bmatrix}
         \matr{H}_0      & \matr{H}_{r-1}      & \cdots & \cdots & \matr{H}_2
                         & \matr{H}_1 \\ 
         \matr{H}_1      & \matr{H}_0          & \cdots & \cdots  & \matr{H}_3
                         & \matr{H}_2 \\ 
         \vdots          & \vdots              & \ddots &             & 
         \vdots          & \vdots \\ 
         \vdots          & \vdots              &        & \ddots & 
         \vdots          & \vdots \\
         \matr{H}_{r-2}   & \matr{H}_{r-3}      & \cdots & \cdots &
         \matr{H}_0      & \matr{H}_{r-1} \\ 
         \matr{H}_{r-1}   & \matr{H}_{r-2} & \cdots & \cdots &
         \matr{H}_1      & \matr{H}_0 
  \end{bmatrix},
\end{align*}
where the scalar $J \times L$-matrices $\matr{H}_0, \matr{H}_1, \ldots,
\matr{H}_{r-1}$ satisfy $\matrHQC{r}(X) = \matr{H}_0 + \matr{H}_1 X + \cdots +
\matr{H}_{r-1} X^{r-1}$.
\begin{Example}\label{cubic:matrices:decomposition}
The code in Example~\ref{cubic:matrices} has
$\matrHQC{3}(X) = \matr{H}_0 + \matr{H}_1 X + \matr{H}_{2} X^{2}$,
where
{\small \begin{align}
   \matr{H}_0
     &\defeq
      \begin{bmatrix}
          1& 1 & 0\\
          0& 1 & 0\\
          0 & 1 & 1
       \end{bmatrix}, 
   \matr{H}_1
      \defeq
      \begin{bmatrix}
          0& 0& 0\\
          0& 0 & 1\\
          0 & 0 & 0
       \end{bmatrix}, 
    \matr{H}_2
      \defeq
       \begin{bmatrix}
          0& 0 & 0\\
          1& 0 & 0\\
          0 & 0 & 0
       \end{bmatrix}.
         \label{matrices:cobiccover:code:1}
\end{align}}
\end{Example}

Given the polynomial parity-check matrix of a QC-code, it is easy to see the
natural connection that exists between quasi-cyclic codes and convolutional
codes (see, e.g., \cite{Tanner:87:1, Levy:Costello:93:1,
  Esmaeili:Gulliver:Secord:Mahmoud:98:1,
  Sridharan:Costello:Sridhara:Fuja:Tanner:02:1, Sridharan:Costello:02:1}).
Briefly, with any QC block code $\codeCQC{r}$ of length $r \cdot L$, given by
a $J\times L$ polynomial matrix parity-check matrix $\matr{H}_{\rm
  QC}^{(r)}(X) = \matr{H}_0 + \matr{H}_1 X + \cdots + \matr{H}_{r-1} X^{r-1}$,
with polynomial operations performed modulo $X^r-1$, we can associate a rate
$(L-J)/L$ convolutional code $\codeCconv$ given by the same $J\times L$
polynomial parity-check matrix $\matrHconv(D) = (h_{jl}(D))_{J\times
  L}=\matrHQC{r}(D) = \matr{H}_0 + \matr{H}_1 D + \cdots + \matr{H}_{r-1}
D^{r-1}$, where the change of variables indicates the lack of modulo $D^r-1$
operations~\cite{Tanner:Sridhara:Sridharan:Fuja:Costello:04:1}. Let the {\em
  syndrome former memory} $\ms\leq r-1$ of $\codeCconv$ be the largest integer in $\{
0, 1, \ldots, r-1 \}$ such that $\matr{H}_{\ms}\neq 0$. Then the polynomial parity-check matrix $\matrHconv(D)$ has
the following scalar description (see, e.g.,
\cite{Johannesson:Zigangirov:99:1})
\begin{align*}
  \matrHbarconv
    &\defeq
       \begin{bmatrix}
         \matr{H}_0    &            &            &      \\
         \matr{H}_1    & \matr{H}_0 &            &      \\
         \matr{H}_2    & \matr{H}_1 & \matr{H}_0 &      \\
         \vdots        & \vdots     & \vdots     & \ddots\\
         \vdots        & \vdots     & \vdots     & \ddots\\
         \matr{H}_{r-1} & \matr{H}_{r-2}       & \matr{H}_{r-3}
         & \ddots \\
                             & \matr{H}_{r-1}   &\matr{H}_{r-2}
         & \ddots \\
         &                   & \matr{H}_{r-1}   & \ddots\\
         &                   &                       & \ddots
       \end{bmatrix}
     = \begin{bmatrix}
         \matr{H}_0    &            &            &      \\
         \matr{H}_1    & \matr{H}_0 &            &      \\
         \matr{H}_2    & \matr{H}_1 & \matr{H}_0 &      \\
         \vdots        & \vdots     & \vdots     & \ddots\\
         \matr{H}_{\ms} & \matr{H}_{\ms-1}       & \matr{H}_{\ms-2}
         & \ddots \\
                             & \matr{H}_{\ms}   &\matr{H}_{\ms-1}
         & \ddots \\
         &                   & \matr{H}_{\ms}   & \ddots\\
         &                   &                       & \ddots
       \end{bmatrix}.
\end{align*}
(Note that $\matrHbarconv$ is a semi-infinite parity-check matrix.) 

\begin{figure}
  \begin{center}
    \epsfig{file=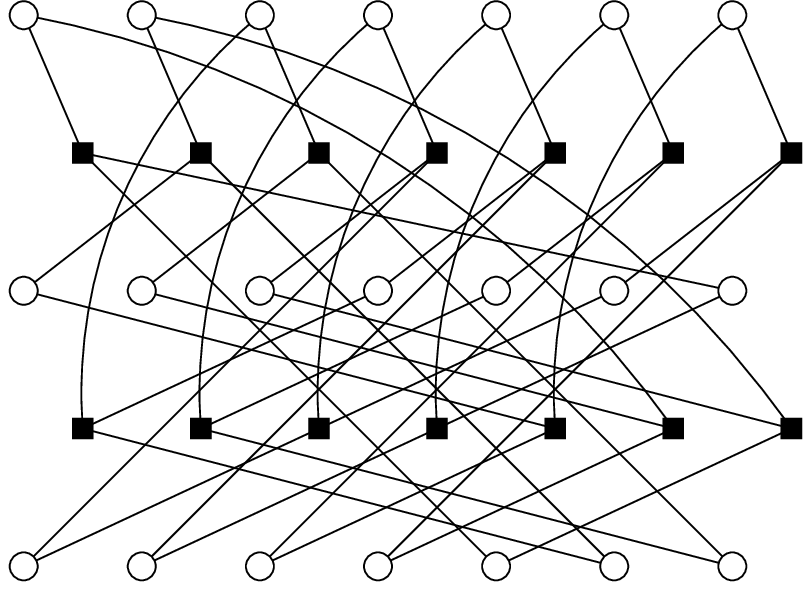,   height=2.5cm}
    \hspace{2cm}
    \epsfig{file=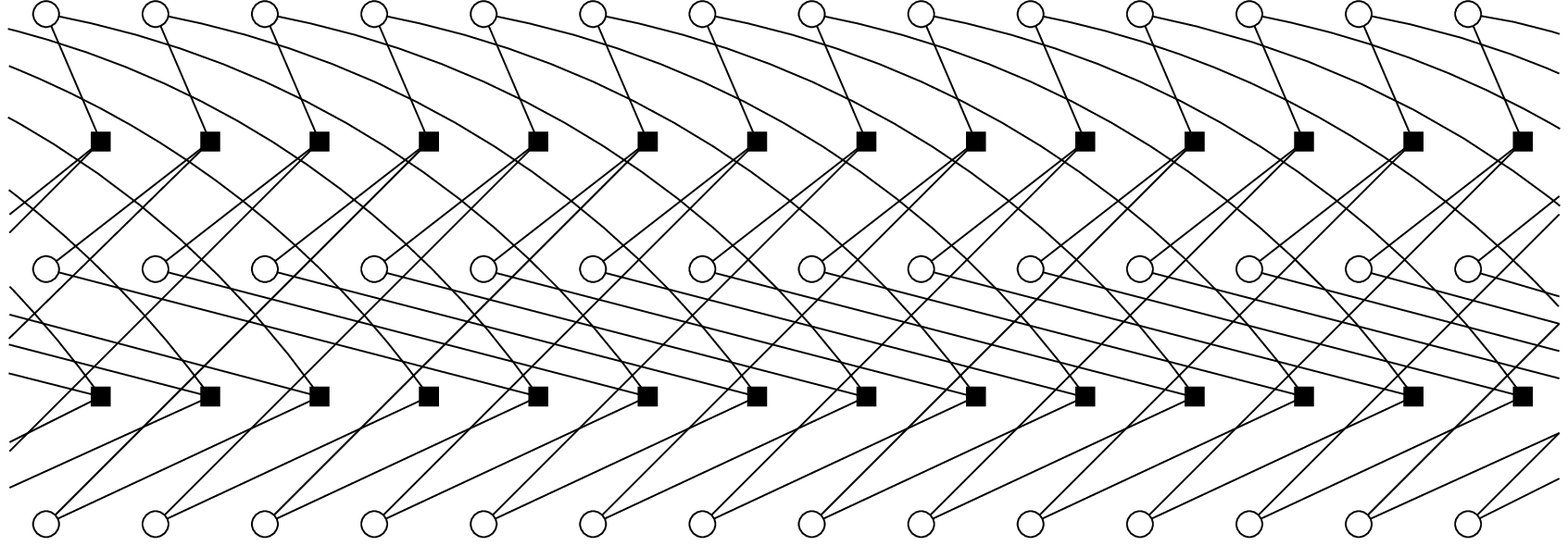, height=2.5cm}
  \end{center} 
  \caption{Left: The finite Tanner graph of a QC code.  Right: The infinite
    Tanner graph of an unwrapped convolutional code.}
  \label{fig:tanner:graph:1}
\end{figure}

\begin{Example}\label{conv:matrix}
  Unwrapping the code in Example~\ref{cubic:matrices}, we obtain a
  convolutional code with polynomial parity-check matrix $\matrHconv$
  given by 
\begin{align}
  \matrHconv(D)=
    & 
       \begin{bmatrix}
          1& 1 & 0\\
          D^2& 1 & D\\
          0  & 1 & 1
       \end{bmatrix}
         \label{eq:pcm:conv:code:1}. 
\end{align}
The syndrome former memory is equal to 2 and the decomposition
leading to the scalar matrix $\matrHbarconv$ is $\matrHconv=
\matr{H}_0 + \matr{H}_1 D + \matr{H}_{2} D^{2},$ where
$\matr{H}_0,\matr{H}_1,\matr{H}_2$ are the same matrices as
in~\eqref{matrices:cobiccover:code:1}.\footnote{Note that this code is
trivially zero.}
\end{Example}
In the same way as we obtained $\codeCconv$ from $\codeCQC{r}$, we
can, upon choosing a positive integer $r$, obtain a code
$\codeCQC{r}(X)$ from a given code $\codeCconv(D)$.\footnote{In this
paper, unless specifically stated otherwise, we will consider the case
$r-1 \geq \ms$, i.e.,~$r \geq \ms+1$.  Some of the results can be
stated for an arbitrary $r$.} It easily follows that, similar to obtaining 
$\matrHbarconv$ by unwrapping $\matrHbarQC{r}$,
$\matrHbarQC{r}$ is obtained from $\matrHbarconv$ by suitably wrapping
$\matrHbarconv$. This unwrapping/wrapping can also be observed in the
Tanner graph representations of $\matrHQC{r}(X)$ and $\matrHconv(D)$,
as illustrated in Fig.~\ref{fig:tanner:graph:1}.

We turn now our attention to codewords in these codes. Codewords in
$\codeCQC{r}$ will be denoted by a polynomial vector like $\vect{c}(X)$
and codewords in $\codeCconv$ will be denoted by a polynomial vector
like $\vect{c}(D)$ (when the code is represented by a scalar
parity-check matrix $\matrHbarconv$, then it will be denoted by a
vector like $\vectbar{c}$). (We will later use a similar notation for
pseudo-codewords.)

For any non-zero codeword $\vect{c}(D)$ with finite support in the
convolutional code, its $r$ {\em wrap-around}, defined by the vector
$\vect{c}(X) \mod (X^r-1) \in \big( \ringFXr \big)^L$, is a codeword
in the associated QC-code, since $\matrHQC{r}(X)\cdot \vect{c}(X) =
\vect{0}$ in $\big( \ringFXr \big)^L$.

In addition, the Hamming weight of the two codewords is linked by the
following inequality $\wH\big(\vect{c}(X) \mod (X^r-1) \big) \leq
\wH(\vect{c}(D))$, which gives the inequality~\cite{Tanner:87:1,
Levy:Costello:93:1} (we assume all codes are non-trivial) 
\begin{align*}
  \dmin\left( \codeCQC{r} \right) 
    &\leq \dfree\left( \codeCconv \right)
          \quad \quad \text{ for all  $r \geq \ms + 1$}.
\end{align*}
Moreover, we have the following inequalities.   
\begin{lemma}
\begin{align*}
  \dmin\left( \codeCQC{r} \right)
    &\leq \dmin\left( \codeCQCtwor \right) 
     \leq \dmin\left( \codeCQCfourr \right)
     \leq \ \cdots \quad \quad \text{for all $r \geq \ms + 1$},
\end{align*}
and
\begin{align*}
  \lim_{r \rightarrow \infty}
    \dmin\left( \codeCQC{r} \right)
      &= \dfree\left( \codeCconv \right).
\end{align*}
\end{lemma}
\begin{pr} Let $\matrHQC{2r}(X)$ be a parity-check matrix of $
  \codeCQCtwor$ and $\matrHQC{r}(X)\defeq\matrHQC{2r}(X)\mod (X^r-1)$ be the
  corresponding parity-check matrix of $\codeCQC{r}$. Let $\vect{c}(X)$ be a
  nonzero codeword in $ \codeCQCtwor$ of weight equal to the minimum distance
  $\dmin\left( \codeCQCtwor \right)$. We have $\matrHQC{2r}(X)\cdot
  \vect{c}(X) = \vect{0}$ in $\big( \ringFXtwor \big)^L$ and since
  $X^{r}-1|X^{2r}-1$, it follows that $\matrHQC{r}(X)\cdot \vect{c}(X) =
  \vect{0}$ in $\big( \ringFXr \big)^L$.
  
  If $\vect{c}(X)\mod (X^r-1)\neq \vect{0}$ then we obtain the inequality
  $\dmin\left( \codeCQC{r} \right) \leq \dmin\left( \codeCQCtwor \right).$ If
  $\vect{c}(X)= \vect{0}\mod (X^r-1)$ we can write $\vect{c}(X)=
  (X^r-1)\vect{c}_1(X)$ with $\vect{c}_1(X)\neq \vect{0}\mod (X^r-1)$
  (otherwise $\vect{c}(X)=\vect{0}\mod (X^{2r}-1)$.)  Hence
  $\matrHQC{r}(X)\cdot (X^r-1)\cdot\vect{c}_1(X) = \vect{0}\mod (X^{2r}-1)$ or
  equivalently, $\matrHQC{r}(X)\cdot \vect{c}_1(X) = \vect{0}\mod (X^{r}-1)$,
  which implies that $ \vect{c}_1(X)$ is a nonzero codeword in $\codeCQC{r}$.
  The Hamming weight $\dmin\left( \codeCQCtwor \right)
  =\wH\big(\vect{c}(X)\big) =\wH\big((X^{r}-1)\vect{c}_1(X)\big)\geq 2\cdot
  \wH(\vect{c}_1(X)\mod (X^r-1))>\wH(\vect{c}_1(X)\mod (X^r-1))\geq
  \dmin\left( \codeCQC{r} \right).$ 
Mimicking this proof we also obtain $\dfree\left( \codeCconv \right)\geq
\dmin\left( \codeCQC{r} \right),$
and hence the desired inequality 
\begin{align*}
  \dmin\left( \codeCQC{r} \right) &\leq \dmin\left( \codeCQCtwor \right) \leq
  \dmin\left( \codeCQCfourr \right) \leq \ \cdots \leq \dfree\left( \codeCconv \right)\quad \quad \text{for all $r
    \geq \ms + 1$}.
\end{align*}
  
From the way we construct the semi-infinite sliding matrix of the convolutional 
by unwrapping the scalar parity-check matrix of the QC block code versions, we
can see that there exists a QC code of circulant size $r$ large enough, so
that its minimum distance is equal to the free distance of the convolutional
code. This assures the limit equality above. 
\end{pr}

If we denote by $\graph{G}\defeq\graph{G}^{(r)}$, $\graph{G}^{(2r)}$,
$\graph{G}^{(4r)}$, $\ldots$ the Tanner graphs associated with the
parity-check matrices of $\codeCQC{r}$, $\codeCQCtwor$, $\codeCQCfourr$,
$\ldots$ of increasing size (which are associated with the same polynomial
matrix $\matr{H}(D)$), it is easy to see that we obtain a {\em tower of
  covers} of the graph $\graph{G}$. The above relationship between the minimum
distances of the codes in the tower is then easily verified in graph language,
since a codeword $\vect{c}(D)$ in a larger graph, say $\codeCQCfourr$, when
projected onto the graphs of $\codeCQC{r}$ and $\codeCQCtwor$ using the
formula $\vect{c}(X)\mod (X^r-1)$, respectively, $\vect{c}(X)\mod (X^{2r}-1)$,
gives another codeword.  Finally, the graph of the associated convolutional
code is an infinite cover of each of the graphs in the tower.

\begin{Example}\label{codewords}
  Using the graphs associated with the
  trivial code $\code{C}=\{(0,0,0)\}$ and its cubic cover
  $\code{C^{(3)}}$ having parity-check matrix $\matrHQC{3}$ 
  (see Example~\ref{cubic:matrices}) corresponding to the codeword $(0,0,0)$ in
  $\code{C}$, we can identify the codeword $\vectbar{c}\defeq
  (1{:}1{:}0, \ 1{:}1{:}0, \ 1{:}1{:}0)\in {\rm Nullsp}(\matrHQC{3})$
  in $\code{C^{(3)}}$. The polynomial
  description of $\vectbar{c}$ is $\vect{c}(X)\defeq(1+X,1+X, 1+X)\in
  {\rm Nullsp}(\matrHQC{3}(X)).$ 
  The wrap-around of $\vect{c}(X)$ $\mod (X-1)$, $\vect{c}(X) \mod (X-1)
  \in\big( \ringFXr \big)^3, r=1$, gives the codeword $(0, 0, 0)$ in
  $\code{C}.$ The associated convolutional code is trivially zero.
\end{Example}

\subsection{The Fundamental Cone of the Parity-Check Matrices of 
                   QC and Convolutional Codes}
 
\label{sec:fundam:cone:par:check:matrix:1}


In this section we introduce our main object of study, the fundamental cone of
a matrix, a set that contains all relevant
pseudo-codewords~\cite{Koetter:Vontobel:03:1, Vontobel:Koetter:05:1:subm,
  Vontobel:Koetter:04:2, Feldman:03:1, Feldman:Wainwright:Karger:05:1,
  Wiberg:96}. 


\begin{definition}[see~\cite{Koetter:Vontobel:03:1, 
  Vontobel:Koetter:05:1:subm,Feldman:03:1, Feldman:Wainwright:Karger:05:1}]
  \label{def:fundamental:cone:1}

  Let $\matr{H}$ be a binary matrix of size $m \times n$, let $\set{I} \defeq
  \set{I}(\matr{H}) \defeq \{ 0, \ldots, n-1 \}$ be the set of column indices
  of $\matr{H}$, and let $\set{J} \defeq \set{J}(\matr{H}) \defeq \{ 0,
  \ldots, m-1 \}$ be the set of row indices of $\matr{H}$, respectively. For
  each $j \in \set{J}$, we let $\set{I}_j \defeq \set{I}_j(\matr{H}) \defeq
  \big\{ i \in \set{I} \ | \ h_{ji} = 1 \big\}$ be the support of the $j$-th
  row of $\matr{H}$. The {\em fundamental polytope} $\fp{P}\defeq \fph{P}{H}$
  of $\matr{H}$ is then defined as
  \cite{Koetter:Vontobel:03:1, Vontobel:Koetter:05:1:subm}
  \begin{align*} 
    \fp{P}
      &\defeq 
         \bigcap_{j=1}^{m} 
           \convhull(\code{C}_j)
         \quad \text{ with } \quad
         \code{C}_j
           \defeq 
             \left\{ \vc \in \{0, 1\}^n 
               \ \left| \ 
               \vect{\vect{r}}_j \vc^\tr = 0 \, \operatorname{mod}\, 2
               \right. 
             \right\},
  \end{align*}
  where $\vect{r}_j$ is the $j$-th row of $\matr{H}$, and
  $\convhull(\code{C}_j)$ is the convex hull of $\code{C}_j$, defined as the
  set of convex combinations of points in $\code{C}_j$ when seen  as
  points in $\R^n$. The {\em fundamental cone} $\fch{K}{H}$ of $\matr{H}$ is
  defined as the conic hull of the fundamental
  polytope $\fp{P}$, which includes the vertex $\vect{0}$, stretched to
  infinity.  Note that if $\vomega \in \fch{K}{H}$, then also $\alpha \cdot
  \vomega \in \fch{K}{H}$ for any real $\alpha > 0$. Moreover, for any
  $\vomega \in \fch{K}{H}$, there exists an $\alpha > 0$ (in fact, a whole
  interval of $\alpha$'s) such that $\alpha \cdot \vomega \in
  \fph{P}{H}$.\edefinition
\end{definition} 


\begin{figure}
  \begin{center}
    \epsfig{file=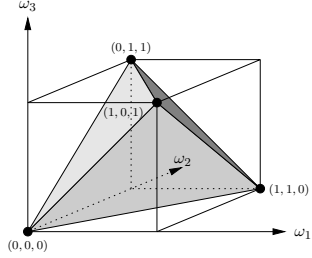, width = 4cm}
  \end{center}
  \caption{The fundamental polytope for 
           $\matr{H} = [ 1 ~ 1 ~ 1 ]$.}
  \label{fig:polytope:1}
\end{figure} 


Vectors in $\fph{P}{H}$ are called {\em pseudo-codewords of $\matr{H}$}, and
we will call any vector in $\fch{K}{H}$ a pseudo-codeword and two
pseudo-codewords that are equal up to a positive scaling constant will be
considered to be equivalent. Clearly, not all pseudo-codewords are codewords,
but all codewords are pseudo-codewords.

The fundamental cone and polytope can both be described by certain sets of
inequalities that are computationally useful~\cite{Vontobel:Koetter:05:1:subm,
  Feldman:Wainwright:Karger:05:1}.  Since our study of pseudo-codewords will
rely heavily on the fundamental cone, we now describe it in more detail.

A vector $ \vomega=(\omega_1, \ldots,\omega_n ) \in \R^n $ is in the
fundamental cone $\fc{K} \defeq \fch{K}{H}$ if and only if
\begin{align}
  \omega_i
     \geq 0
  \quad 
  & \text{for all  $i \in \set{I}(\matr{H})$},
        \label{eq:fc:inequalities:1:1} \\
  \sum_{i \in (\set{I}_j \setminus \{ i' \})} 
    \omega_i 
    \geq \omega_{i'} 
  \quad 
  & \text{for all $j \in \set{J}(\matr{H})$ and 
          for all $i' \in \set{I}_j(\matr{H})$}.
    \label{eq:fc:inequalities:1:2}
\end{align}


\begin{Example} 
  Let $\matr{H} \defeq [1, \ 1, \ 1]$ be a parity-check matrix for the code
  $\code{C} = \{ (0, 0, 0), \ (1, 1, 0), $ $(1, 0, 1), \ (0, 1, 1) \}$. The
  fundamental polytope $\fph{P}{H}$ associated with $\matr{H}$ is (see
  Fig.~\ref{fig:polytope:1})
  \begin{align*}
    \fph{P}{H}
      &= \left\{
           (\omega_1, \omega_2, \omega_3)\in \R^3 
           \left|
           \begin{array}{l}
             0 \leq \omega_1 \leq 1 \\
             0 \leq \omega_2 \leq 1 \\
             0 \leq \omega_3 \leq 1
           \end{array}
           \quad
           \begin{array}{l}
             - \omega_1 + \omega_2 + \omega_3 \geq 0 \\
             + \omega_1 - \omega_2 + \omega_3 \geq 0 \\
             + \omega_1 + \omega_2 - \omega_3 \geq 0 \\
             + \omega_1 + \omega_2 + \omega_3 \leq 2
           \end{array}
           \right.
         \right\},
  \end{align*}
  from which we obtain the fundamental cone $\fph{K}{H}$ associated with
  $\matr{H}$:
  \begin{align*}
    \fph{K}{H}
      &= \left\{
           (\omega_1, \omega_2, \omega_3)\in \R^3 
           \left|
           \begin{array}{l}
             0 \leq \omega_1 \\
             0 \leq \omega_2 \\
             0 \leq \omega_3
           \end{array}
           \quad
           \begin{array}{l}
             - \omega_1 + \omega_2 + \omega_3 \geq 0 \\
             + \omega_1 - \omega_2 + \omega_3 \geq 0 \\
             + \omega_1 + \omega_2 - \omega_3 \geq 0
           \end{array}
           \right.
         \right\}.
  \end{align*}
  We see that the pseudo-codeword $\vomega = \bigl( \frac{2}{3}, \frac{2}{3},
  \frac{2}{3} \bigr)$ satisfies the inequalities of both the fundamental
  polytope and the fundamental cone. Also, the pseudo-codeword $\vomega = (1,
  1, 1)$ and its positive multiples are in the fundamental cone but not
  necessarily in the fundamental polytope.
\end{Example}


\begin{remark}
  As explained briefly in Sec.~\ref{sec:introduction}, pseudo-codewords can
  also be described using graph theory language as vectors corresponding
  to codewords in a cover graph.  Looking again at the cover tower $\graph{G}
  \defeq \graph{G}^{(r)}$, $\graph{G}^{(2r)}$, $\graph{G}^{(4r)}$, $\ldots$, a
  codeword $\vect{c}(D)$ of a larger graph, say $\codeCQCfourr$, seen as a
  vector with real components, when projected over $\R$ onto the graphs of
  $\codeCQC{r}$ and $\codeCQCtwor$, using the formula $\vect{c}(X)\mod
  (X^r-1)$, respectively, $\vect{c}(X)\mod (X^{2r}-1)$, gives a
  pseudo-codeword. Similarly, a codeword $\vect{c}(D)$ of the unwrapped
  convolutional code $\codeCconv$, when projected onto the graphs of
  $\codeCQC{r}$ $\codeCQCtwor$, $\codeCQCfourr, \ldots,$ using the formulas
  $\vect{c}(X)\mod (X^r-1)$, $\vect{c}(X)\mod (X^{2r}-1)$, $\vect{c}(X)\mod
  (X^{4r}-1), \ldots$, gives pseudo-codewords in the QC codes. 
\end{remark}

\begin{Example}
  Let $\vectbar{c}\defeq (1{:}1{:}0, \ 1{:}1{:}0, \ 1{:}1{:}0)\in {\rm
    Nullsp}(\matrHQC{3})$ be the codeword identified in
  Example~\ref{codewords} for the code $\code{C^{(3)}}$, with corresponding
  polynomial description $\vect{c}(X)\defeq(1+X,1+X, 1+X)\in {\rm
    Nullsp}(\matrHQC{3}(X))$.  When projecting $\vect{c}(X)$ over $\R$ onto
  the graph of $\code{C}=\{(0,0,0)\}$ we obtain $\vect{c}(X)\mod
  (X-1)=(2,2,2)$, which is a pseudo-codeword in $\code{C}$, equivalent to the
  pseudo-codeword $\vomega = \bigl( \frac{2}{3}, \frac{2}{3}, \frac{2}{3}
  \bigr)$ presented in Example~\ref{example:introduction}. The fundamental
  cone of the convolutional code given by $\matrHconv(D)$ in
  Example~\ref{conv:matrix} is trivially zero, i.e., it does not contain any
  non-zero pseudo-codewords.

\end{Example}
\begin{remark}
  In the above example we made the following important connection between two
  graphical ways of representing pseudo-codewords $\vomega$ of a QC code
  $\code{C}$ of length $rL$ as vectors corresponding to codewords
  $\vect{c}(X)$ in some QC code (or convolutional code)  associated with an
  $m$-cover (or an infinite cover) of the graph of $\code{C}$. We can see them
  as vectors with each entry representing an average of the ones attributed to
  the variable nodes of the cover graph that are in the ``cloud'' of that
  entry, e.~g., $\vomega = \bigl( \frac{2}{3}, \frac{2}{3}, \frac{2}{3}
  \bigr)$, corresponding to pseudo-codewords in the polytope of $\code{C}$, or
  we can see them as vectors that are the projections of $\vect{c}(X)$ onto
  the code $\code{C}$ by the formula $\vect{c}(X)\mod (X^r-1)$, e.g., $\vomega
  = ( 2, 2, 2)$, corresponding to pseudo-codewords in the fundamental cone of
  $\code{C}$ that are not necessarily in the polytope.
  
  In fact a similar statement can be made about pseudo-codewords in any block
  code $\code{C}$ of length $rL$, not having necessarily a QC structure. A
  pseudo-codeword $\vomega$ in $\code{C}$ is a vector corresponding to some
  codeword $\vectbar{c}$ in some block code (or convolutional code) associated
  with an $m$-cover (or an infinite cover) of the graph of $\code{C}$ with
  each entry representing an average of the ones attributed to the variable
  nodes of the cover graph that are in the ``cloud'' of that entry. However,
  this is the same, under the equivalence between different representations of
  a pseudo-codeword in the fundamental cone, as defining a pseudo-codeword
  $\vomega$ as a vector for which there exists a cover corresponding to a
  block code or to a convolutional code, $\codebar{C}$, of the code
  $\code{C}$, and there exists some codeword, $\vectbar{c}$ in $\codebar{C}$,
  such that $\vomega= \tilde{\vect{c}} \mod (rL)$ where the vector $
  \tilde{\vect{c}}$ is equal to a permutation of the vector $\vectbar{c}$ (where
  the components of $\vectbar{c}$ are taken in the following
  order  $1, m+1, \ldots, (L-1)m+1, 2, m+2, \ldots, (L-1)m+2, \ldots, m-1,
  2m-1, \ldots, Lm-1$).  The vector $\tilde{\vect{c}}$ it is seen as a vector
  with real entries and the modular operation $\mod (rL)$ is performed over the
  real numbers.
\end{remark}

The above description of the fundamental polytope and cone provides
computationally easy descriptions of the pseudo-codewords, especially in the
case of QC and convolutional codes.
From~\eqref{eq:fc:inequalities:1:1} and~\eqref{eq:fc:inequalities:1:2} it
follows that for any parity-check matrix $\matr{H}$ there is a matrix
$\matr{K}$ such that a vector $\vomega$ is a pseudo-codeword in the
fundamental code $\fc{K}(\matr{H})$ if and only if $\matr{K} \cdot \vomega^\tr
\geq \vect{0}^\tr$. In the case of QC and convolutional codes, the fundamental
cone can be described with the help of polynomial matrices. Namely, for any
$\matrHQC{r}(X)$ there is a polynomial matrix $\matrKQC{r}(X)$ such that
$\vomega(X) \in \fc{K}\bigl( \matrHQC{r}(X) \bigr)$ if and only if
$\matrKQC{r}(X) \cdot \vomega(X)^\tr \geq \vect{0}^\tr$. Similarly, for any
$\matrHconv(D)$ there is a polynomial matrix $\matrKconv(D)$ such that
$\vomega(D) \in \fc{K}\bigl( \matrHconv(D) \bigr)$ if and only if
$\matrKconv(D) \cdot \vomega(D)^\tr \geq \vect{0}^\tr$.



This fundamental-cone description is particularly simple and hence useful in
the case of {\bf monomial} parity-check matrices $\matrHconv$
(see~\cite{Tanner:Sridhara:Sridharan:Fuja:Costello:04:1}), since the
fundamental-cone inequalities can then be stated very simply as follows. A
vector $\vomega$ is a pseudo-codeword in the fundamental cone of a monomial
matrix $\matr{H_{\rm conv}}$ if and only if the associated polynomial vector
$\vomega(D) = \bigl( \omega_1(D), \omega_2(D), \ldots, \omega_L(D) \bigr)$
satisfies:
\begin{align*}
  \omega_{\ell}(D) 
    \geq 0, 
    &\quad \text{for all $\ell \in \{ 1, \ldots, L \}$}, \\
  \sum_{\ell \in \{1,\ldots,L\} \setminus \{ \ell' \}}
    h_{j\ell}(D) \omega_{\ell}(D) \geq h_{j\ell'}(D) \omega_{\ell'}(D), 
    &\quad \text{for all $j \in \{1, \ldots, J \}$, 
                 for all $\ell'\in \{1,\ldots, L\}$}.
\end{align*}
Equivalently, if $\matrOmega(D)$ is an $L\times L$ matrix with $\ell$-th row
vector 
\begin{align*}
  \begin{pmatrix}
    +\omega_1(D)     & \cdots & +\omega_{\ell-1}(D) & -\omega_{\ell}(D) &
    +\omega_{\ell+1}(D) & \cdots & \omega_L(D)
  \end{pmatrix}
\end{align*}
for $j \in \{ 1, \ldots, L \}$, then $\vomega(D)$ is a pseudo-codeword if and
only if $\vomega_{\ell}(D) \geq \vect{0}$ for all $\ell \in \{ 1, \ldots, L
\}$ and
\begin{align*}
  \matrHconv(D)
  \cdot
  \matrOmega(D)^\tr
    &\geq \matr{0}.
\end{align*}


\begin{Example}
  \label{ex:smallexample:1} 

  Let 
  \begin{align*}
    \matrHconv(D)
      &\defeq
         \begin{bmatrix}
           1 & 1     & 1      & 1 \\
           1 & D     & D^{2}  & D^3 \\
           1 & D^{4} & D^3    & D^{2}
         \end{bmatrix}
  \end{align*}
  be a polynomial parity-check matrix of a rate-$1/4$ convolutional code. The
  following vector
  \begin{align*}
    \vomega(D)
      &\defeq
         \begin{pmatrix}
           3D^2+D^3 & 4D+D^2 & 3+D+4D^2+D^3 & 3+4D+D^2
         \end{pmatrix} \\
      &= \begin{pmatrix}
           0 & 0 & 3 & 3
         \end{pmatrix}
         +
         \begin{pmatrix}
           0 & 4 & 1 & 4
         \end{pmatrix}
         D 
         + 
         \begin{pmatrix}
           3 & 1 & 4 & 1
         \end{pmatrix}
         D^2
         +
         \begin{pmatrix}
           1 & 0 & 1 & 0
         \end{pmatrix}
         D^3,
  \end{align*}
  corresponding to the scalar vector
  \begin{align*}
    \vomegabar
      &= \left(
             0, 0, 3, 3, \
             0, 4, 1, 4, \
             3, 1, 4, 1, \
             1, 0, 1, 0, \
             \cdots
         \right),
  \end{align*}
  is a pseudo-codeword for the convolutional code since $\vomega(D) \geq
  \vect{0}$ and the matrix
  \begin{align*}
    \matrOmega(D)\defeq
    \begin{bmatrix}
      -3D^2-D^3 & +4D+D^2 & +3+D+4D^2+D^3 & +3+4D+D^2 \\
      +3D^2+D^3 & -4D-D^2 & +3+D+4D^2+D^3 & +3+4D+D^2 \\
      +3D^2+D^3 & +4D+D^2 & -3-D-4D^2-D^3 & +3+4D+D^2 \\
      +3D^2+D^3 & +4D+D^2 & +3+D+4D^2+D^3 & -3-4D-D^2
    \end{bmatrix}
  \end{align*}
  gives
  {\small
  \begin{align*}
    \matrHconv(D)
    \cdot
    \matrOmega(D)^\tr
      &= \begin{bmatrix}
           6+9D+3D^2     & 4D^2+4D^3+8D^4+2D^5 & 6D^3+2D^4+8D^5+2D^6 \\
           6+D+D^2+2D^3  & 2D^2+4D^3+8D^4+2D^5 & 6D^2+8D^3+2D^4+2D^6 \\
           7D+D^2        & 4D^2+4D^3           & 6D^2+2D^3           \\
           D+7D^2+ 2D^3  & 10D^2               & 8D^5+2D^6 
         \end{bmatrix}^\tr \\
      &\geq \matr{0},
  \end{align*}%
  }
  i.e., $\matrHconv(D) \cdot \matrOmega(D)^\tr$ has all polynomial
  entries with non-negative coefficients.
\end{Example}


Similarly, for a QC code that is described by a {\bf monomial} $J \times L$
parity-check matrix $\matrHQC{r}(X)$, a vector $\vomega(X)$ is a
pseudo-codeword in the corresponding fundamental cone if and only if the
associated polynomial vector $\vomega(X) = \bigl( \omega_1(X), \omega_2(X),
\ldots, \omega_L(X) \bigr)$ satisfies
\begin{align*}
  \big( \omega_{\ell}(X) \mod(X^r-1) \big) \geq 0 
    &\quad \quad \text{for all $\ell \in \{ 1, \ldots, L \}$},
\end{align*}
and
\begin{align*}
  \left(\matrHQC{r}(X) \matrOmega(X)^\tr \mod(X^r-1)\right)
    \geq \matr{0}.
\end{align*}


\begin{Example}
  \label{ex:small:example:QC:1}  

  Let $\matrHQC{5}(X)$ be the $r=5$ parity-check matrix obtained from
  $\matrHconv(D)$ in Example~\ref{ex:smallexample:1} for a QC block code of
  length $n=20$. Then, for the polynomial vector
  \begin{align*}
    \vomega^{(5)}(X)
      &\defeq
         \begin{pmatrix}
           3X^2+X^3 & 4X+X^2 & 3+X+4X^2+X^3 & 3+4X+X^2
         \end{pmatrix},
  \end{align*}
  we obtain
  \begin{align*}
    &
    \matrHQC{5}(X)\matrOmega(X)^\tr\mod(X^5-1) \\
      &\quad\quad
       = \begin{bmatrix}
           6+9X+3X^2     & 2+4X^2+4X^3+8X^4 & 8+2X+6X^3+2X^4 \\
           6+X+X^2+2X^3  & 2+2X^2+4X^3+8X^4 & 2X+6X^2+8X^3+2X^4 \\
           7X+X^2        & 4X^2+4X^3        & 6X^2+2X^3           \\
           X+7X^2+ 2X^3  & 10X^2            & 8+2X
         \end{bmatrix}^\tr.
  \end{align*}
  Since $\vomega^{(5)}(X) \geq \vect{0}$ and $(\matr{H}_{\rm
    QC}^{(5)}(X)\matrOmega(X)^\tr\mod(X^5-1)) \geq \matr{0}$, we know that
  $\vomega^{(5)}(X)$ is a pseudo-codeword.
\end{Example}


The pseudo-codeword $\vomega(D) = (3D^2+D^3, \ 4D+D^2, \ 3+D+4D^2+D^3, \ 
3+4D+D^2)$ in Example~\ref{ex:smallexample:1} was obtained by using a method
to be described in
Sec.~\ref{sec:analysis:problematic:pseudo:codewords:conv:codes:1}.  Projecting
this pseudo-codeword in the convolutional code onto the $r$-wrapped QC code,
for $r=5$, we obtained the pseudo-codeword $\vomega(X) = (3X^2+X^3, \ 4X+X^2,
\ 3+X+4X^2+X^3, \ 3+4X+X^2)$ in the QC code. This is not a mere coincidence as
the following lemma shows.


\begin{lemma}
  \label{lemma:qc:vs:conv:code:pseudo:codeword:1}
  
  Let $\vomega(D)$ be a pseudo-codeword in the convolutional code defined by
  $\matrHconv(D)$, i.e., $\vomega(D) \in \fc{K}(\matrHconv(D))$.  Then its $r$
  wrap-around polynomial vector is a pseudo-codeword in the associated QC-code
  defined by $\matrHQC{r}(X)$, i.e., $\vomega(X) \mod (X^r-1) \in
  \fc{K}(\matrHQC{r}(X))$.
\end{lemma} 


\begin{pr} 
  We have seen that for any $\matrHconv(D)$ describing a convolutional code
  there is a matrix $\matrKconv(D)$ such that a polynomial vector $\vomega(D)$
  is a pseudo-codeword in the fundamental cone $\fc{K}(\matrHconv(D))$ if and
  only if $\matrKconv(D) \vomega(D)^\tr \geq \vect{0}^\tr$.  By reducing
  $\matrKconv(D)$ modulo $D^r-1$, we obtain a matrix $\matrKQC{r}(X) =
  \left(\matrKconv(X) \mod(X^r-1) \right)$ with the property that a polynomial
  vector $\vomega(X)$ is a pseudo-codeword in the fundamental cone
  $\fc{K}(\matrHQC{r}(X))$ if and only if $\matrKQC{r}(X) \vomega(X)^\tr \geq
  \vect{0}^\tr$. Reducing $\matrKconv(D) \vomega(D)^\tr \geq \vect{0}^\tr$
  modulo $D^r-1$, we obtain $\matrKQC{r}(X)(\vomega(X)\mod(X^r-1)) ^\tr \geq
  \vect{0}^\tr$, which proves the claim.
\end{pr} 


\begin{remark}
  This result can be easily deduced using the graph-theory language at the end
  of Sec.~\ref{sec:link:qc:and:conv:codes:1}. Indeed, looking again at the
  cover tower $\graph{G}\defeq\graph{G}^{(r)}, \graph{G}^{(2r)},
  \graph{G}^{(4r)}, \ldots $, a pseudo-codeword $\vect{\vomega}(D)$ of a
  larger graph, say $\codeCQCfourr$, when projected onto the graphs of
  $\codeCQC{r}$ and $\codeCQCtwor$, using the formula $\vect{\vomega}(X)\mod
  (X^r-1)$, respectively, $\vect{\vomega}(X)\mod (X^{2r}-1)$, gives another
  pseudo-codeword. Similarly, a pseudo-codeword $\vect{\vomega}(D)$ of the
  unwrapped convolutional code $\codeCconv$, when projected onto the graphs of
  $\codeCQC{r}$ $\codeCQCtwor$, $\codeCQCfourr, \ldots,$ using the formulas
  $\vect{\vomega}(X)\mod (X^r-1)$, $\vect{\vomega}(X)\mod (X^{2r}-1)$,
  $\vect{\vomega}(X)\mod (X^{4r}-1), \ldots, $ gives pseudo-codewords in the
  QC codes.
\end{remark}

\begin{Example}\label{projecting:conv:pseudo-codewords}
The  pseudo-codeword 
  \begin{align*}
    \vomega(D)
      &\defeq
         \begin{pmatrix}
           3D^2+D^3 & 4D+D^2 & 3+D+4D^2+D^3 & 3+4D+D^2
         \end{pmatrix} 
  \end{align*}
 of the convolutional code in Example~\ref{ex:smallexample:1} projects
onto the following pseudo-codewords in the QC wrapped block code
versions of parity-check matrix $\matrHQC{r}(X)=\matrHconv(X)\mod
(X^r-1)$, for all $r\geq 0 $:\footnote{If the convolutional code has a
monomial polynomial parity-check matrix then we do not need the earlier assumption $r\geq 
 \ms+1$, as  Lemma~\ref{lemma:qc:vs:conv:code:pseudo:codeword:1} holds for each $r\geq 1$.}
\begin{align*}
\vomega^{(r)}(X)
     & \defeq  \vomega(D)\mod (X^{r}-1) =
       \begin{pmatrix}
        3X^2+X^3 & 4X+X^2 & 3+X+4X^2+X^3 & 3+4X+X^2
       \end{pmatrix},\\& {\rm ~for~ all ~ } r\geq 4, ~( r=5 {\rm ~case~above~ is~ included},)  \\
  \vomega^{(3)}(X)
     & \defeq  \vomega(D)\mod (X^{3}-1) =
     \begin{pmatrix}
       1+3X^2 & 4X+X^2 & 4+X+4X^2 & 3+4X+X^2
     \end{pmatrix},\\
   \vomega^{(2)}(X)
     & \defeq \vomega(D)\mod (X^{2}-1 ) =
     \begin{pmatrix}
        3+X & 1+ 4X & 7+2X & 4+4X 
      \end{pmatrix},\\
    \vomega^{(1)}(X)
     & \defeq  \vomega(D)\mod (X-1) ~= 
      \begin{pmatrix} 4& 5 & 9 & 8       
      \end{pmatrix}.
\end{align*}
\end{Example}

\section{Pseudo-Weight Comparison Between 
            QC Codes and Convolutional Codes}

\label{sec:pseudo:weight:comparison:1}


This section begins with an introductory subsection in which various channel
pseudo-weights are defined and continues with the main result that the minimum
AWGNC, BEC, BSC, and max-fractional pseudo-weights of a convolutional code are
at least as large as the corresponding pseudo-weights of a wrapped QC block
code.


\subsection{Definitions of Channel Pseudo-Weights}

\label{sec:pseudo:weight:definitions:1}


\begin{definition}{\!\!\cite{Koetter:Vontobel:03:1,
      Vontobel:Koetter:05:1:subm, Feldman:Wainwright:Karger:05:1,
      Feldman:03:1, Wiberg:96, Forney:Koetter:Kschischang:Reznik:01:1}}
  Let $\vomega = (\omega_0,$ $\ldots, \omega_{n-1})$ be a \emph{nonzero}
  vector in $\Rp^n$. The AWGNC pseudo-weight and the BEC pseudo-weight of the
  vector $\vomega$ are defined to be, respectively,
  \begin{align*}
    \wpsAWGNC(\vomega)
      &\defeq \frac{\onenorm{\vomega}^2}
                   {\twonorm{\vomega}^2},
    \quad\quad
    \wpsBEC(\vomega)
       = \big| \supp(\vomega) \big|,
  \end{align*}
  where $\onenorm{\vomega}$ and $\twonorm{\vomega}$ are the $1$-norm,
  respectively $2$-norm, of $\vomega$, and $\supp(\vomega)$ is the set of all
  indices $i$ corresponding to nonzero components $\omega_i$ of
  $\vomega$. In order to define the BSC pseudo-weight $\wpsBSC(\vomega)$, we
  let $\vomega'$ be the vector of length $n$ with the same components as
  $\vomega$ but in non-increasing order. Now let
  \begin{align*}
    f(\xi)
      &\defeq \omega'_i \quad (i < \xi \leq i+1,\ 0 < \xi \leq n),
     \quad\quad 
    F(\xi)
      \defeq \int_{0}^{\xi} f(\xi') \dint{\xi'}, 
    \quad \text{ and } \quad
    e
       \defeq F^{-1} \left( \frac{F(n)}{2} \right).
  \end{align*}
  Then the binary symmetric channel (BSC)-pseudo-weight $\wpsBSC(\vomega)$ is
  $\wpsBSC(\vomega) \defeq 2e$. Finally, the fractional weight of a vector
  $\vomega \in [0, 1]^n$ and the max-fractional weights of a vector $\vomega
  \in \Rp^n$ are defined to be, respectively,
  \begin{align*}
    \wfr(\vomega)
      &= \onenorm{\vomega},
    \quad\quad
    \wmaxfr(\vomega)
       \defeq
         \frac{\onenorm{\vomega}}
              {\infnorm{\vomega}}, 
  \end{align*}
  where $\infnorm{\vomega}$ is the infinite or max norm. For $\vomega =
  \vect{0}$ we define all of the above pseudo-weights, fractional weights, and
  max-fractional weights to be zero. \edefinition
\end{definition}

A discussion of the motivation and significance of these definitions
can be found in~\cite{Vontobel:Koetter:05:1:subm}. Note that whereas the
fractional weight has an operational meaning only for vertices of the
fundamental polytope, the other measures have an operational meaning for any
vector in the fundamental polytope or cone. Note also that here $\wfr$ and
$\wmaxfr$ are defined for any vector in $\Rp^n$, whereas $\wfr$ and $\wmaxfr$
in~\cite{Feldman:03:1} are what we will call $\wfrmin$ and $\wmaxfrmin$.


\begin{Example} 
  Let
  \begin{align*}
    \vomega(D)
      &= \begin{pmatrix}
           3D^2+D^3 & 4D+D^2 & 3+D+4D^2+D^3 & 3+4D+D^2
         \end{pmatrix}
  \end{align*}
  be the pseudo-codeword in Example~\ref{ex:smallexample:1} and let
  $\vomegabar = (0, 0, 3, 3, \ 0, 4, 1, 4, \ 3, 1, 4, 1, \ 1, 0, 1, 0, \ $ $0,
  0, 0, 0, \ \ldots)$ be its scalar vector description. Then
  \begin{align*}
    \wpsAWGNC(\vomega)
      &= \frac{\onenorm{\vomega}^2} 
              {\twonorm{\vomega}^2}
       = \frac{(3\cdot 4 + 3\cdot 3 + 5\cdot1)^2}
              {3\cdot 16 + 3\cdot 9 + 5\cdot1}
       =8.45,
    \quad\quad
    \wpsBEC(\vomega)
       = |\supp(\vomega)|
       = 11.
  \end{align*}
  In order to compute $\wpsBSC(\vomega)$, we let $\vomegabar' = (4, 4, 4, 3,
  3, 3, 1, 1, 1, 1, 1, 0, 0, 0, 0, 0, \ldots )$ be the vector that lists the
  components of $\vomega$ in non-increasing order. We obtain
  $\wpsBSC(\vomega)=2e=\frac{20}{3} = 6.67$, since we need to add up $e =
  \frac{10}{3}$ ordered components of $\vomega'$ to obtain
  $\frac{\onenorm{\vomega}}{2} = 13$. Finally $\infnorm{\vomega} =
  \max_{i=1}^{16} \omega_i$, from which it follows that $\wmaxfr(\vomega) =
  \frac{\onenorm{\vomega}} {\infnorm{\vomega}} = \frac{26}{4} = 6.5$.
\end{Example}


A measure of the effect that the pseudo-codewords have on the performance of
a code is given by the minimum
\emph{pseudo-weight}~\cite{Koetter:Vontobel:03:1, Vontobel:Koetter:05:1:subm,
  Feldman:03:1, Feldman:Wainwright:Karger:05:1}
\begin{align*}
  \wpsminh{H}
    &\defeq
       \min_{\vomega \in \setV(\fph{P}{H}) \setminus \{ \vect{0} \}}
         \wps(\vomega),
\end{align*}
where $\setV(\fph{P}{H}) \setminus \{ \vect{0} \}$ is the set of all non-zero
vertices in the fundamental polytope $\fph{P}{H}$ and the pseudo-weights are
the appropriate ones for each channel (AWGNC, BSC, and BEC pseudo-weights) and
the minimum fractional and max-fractional weights.

Computing these values can be quite challenging, since the task of finding the
set of vertices of $\fph{P}{H}$ is in general very complex. However, in the
case of four of the above pseudo-weights (the minimum AWGNC, BSC, and BEC
pseudo-weights and the minimum max-fractional weight) there is a
computationally simpler description given by
\begin{align*}
  \wpsminh{H}
    &= \min_{\vomega \in \fch{K}{H} \setminus \{ \vect{0} \}}
         \wps(\vomega),
\end{align*}
for the appropriate pseudo-weight. (Note that there is no such
statement for the minimum fractional weight; see,
e.g.,~\cite{Feldman:Wainwright:Karger:05:1, Vontobel:Koetter:05:1:subm}).


\begin{Example} 
  Let $\matrHQC{5}(X)$ be the matrix of Example~\ref{ex:small:example:QC:1}
  and let $\codeCQC{5}$ be the QC block code of length $n = 20$, with $r = 5$.
  The minimum distance is 6 and it is equal to the $\wpsAWGNCmin(\matr{H}_{\rm
    QC}^{(5)})$. (This was obtained using a vertex enumeration program for
  polytopes that lists all the minimal pseudo-codewords of a code,
  see~\cite{Avis:00:1}.)
\end{Example}

 


\begin{remark}
  \label{remark:relationship:minium:weights:1}

  In~\cite{Vontobel:Koetter:05:1:subm} it was shown that for any code defined
  by a parity-check matrix $\matr{H}$, the following inequalities hold:
  \begin{alignat*}{2}
    \wfrminh{H}
      &\leq
         \wmaxfrminh{H}
      &\leq
         \wpsAWGNCminh{H}
      &\leq
         \wpsBECminh{H}, \\
    \wfrminh{H}
      &\leq
         \wmaxfrminh{H}
      &\leq
         \wpsBSCminh{H}
      &\leq
         \wpsBECminh{H}.
  \end{alignat*}
  Therefore, $\wfrminh{H}$ and $\wmaxfrminh{H}$ can serve as lower bounds 
  for $\wpsAWGNCminh{H}$, $\wpsBSCminh{H}$, and $\wpsBECminh{H}$.
\end{remark}


\subsection{Minimum Pseudo-Weights}
\label{sec:pseudo:weight:inequality:1}


In what follows, we compare the minimum pseudo-weights and minimum
max-fractional weight of a QC block code to those of its corresponding
convolutional code which we assume to have a fundamental cone
containing non-zero vectors. In order to analyze the minimum
pseudo-weight and minimum max-fractional weight, it is sufficient to
analyze the weights of the non-zero vectors in the fundamental
cone. Throughout this section, without loss of generality, all
pseudo-codewords $\vomega(D)$ are assumed to have finite
support.\footnote{With suitable modifications, this can easily be
generalized to $\vomega(D)$ with $\onenorm{\vomega(D)} < \infty$. Note
that such polynomial vectors also fulfill $\twonorm{\vomega(D)} <
\infty$.}


\begin{theorem}
  \label{th:qc:vs:conv:code:pseudo:weight:1}

  For the AWGNC, BEC, and BSC pseudo-weights, if $\vomega(D) \in
  \fc{K}(\matrHconv(D))$, then
  \begin{align*}
    \wps\big(
      \vomega(X) \mod (X^r-1) \big)
      &\leq \wps\big( \vomega(D) \big).
  \end{align*}
  Therefore, if the fundamental cone of the convolutional code is not
  trivial (i.e., it contains non-zero vectors) we obtain
  \begin{align*}
    \wpsmin\left( \matrHQC{r}(X) \right)
      &\leq \wpsmin\big( \matrHconv(D) \big).
  \end{align*}
\end{theorem} 


\begin{pr}
  In the following, we need to analyze separately the AWGNC, BEC, and BSC
  pseudo-weights of $\vomega(D)$ and of its $r$ wrap-around $\vomega(X) \mod
  (X^r-1)$. Let $\vomega(D) = \bigl( \omega_1(D), \ldots, \omega_{L}(D)
  \bigr)$ be a pseudo-codeword. By assumption $\vomega(D)$ has finite support,
  i.e.,~there exists an integer $t$ such that the maximal degree of any
  $\omega_{\ell}(D)$, $\ell \in \{ 1, \ldots, L \}$, is smaller than $t$.


  {\bf Case 1 (AWGNC):} Since $\onenorm{\vomega(D)}= \onenorm{\vomega(X) \mod
    (X^r-1)}$ and
  \begin{align*}
    \twonormbig{\vomega(X) \mod (X^r-1)}^2
      &= \sum_{\ell=1}^{L} 
           \sum_{i=0}^{r-1}
             \left(\sum_{i'=0}^{\lfloor (t-1)/r \rfloor}
               \omega_{\ell, i+i'r}\right)^2
       \geq
         \sum_{\ell=1}^{L}
           \sum_{i=0}^{r-1}
              \sum_{i'=0}^{\lfloor (t-1)/r \rfloor} 
                \omega_{\ell, i+i'r}^2 
       = \twonormbig{\vomega(D)}^2,
  \end{align*}
  we obtain $ \wpsAWGNC(\vomega(D)) \geq \wpsAWGNC(\vomega(X) \mod (X^r-1))$.


  {\bf Case 2 (BEC):} Since the components of the vector $\vomega(X) \mod
  (X^r-1)$ are obtained by adding in $\Rp$ certain non-negative components of
  $\vomega(D)$, it follows that 
  \begin{align*}
    |\supp\vomega(D)|
      &\geq |\supp\left(\vomega(X) \mod (X^r-1)\right)|
  \end{align*}
  and we obtain $\wpsBEC\big( \vomega(D) \big) \geq \wpsBEC\big( \vomega(X)
  \mod (X^r-1) \big)$.


  {\bf Case 3 (BSC):} In order to compare the BSC-pseudo-weight of the two
  vectors, we first need to arrange the components in decreasing order. Let
  $M_0 \geq M_1 \geq \ldots \geq M_{tL-1}$ and $m_0 \geq m_1 \geq \ldots
  \geq m_{rL-1}$ be listings of all the coefficients of all the components of
  $\vomega(D)$ and $\vomega(X) \mod (X^r-1)$, respectively, in non-increasing
  order. Since $\onenorm{\vomega(D)} = \onenorm{\vomega(X) \mod (X^r-1)}$, we
  obtain that $\frac{\onenorm{\vomega(D)}}{2} = \frac{\onenorm{\vomega(X) \mod
      (X^r-1)}}{2}\defeq M$, which gives $\sum_{i=0}^{tL-1} M_i =
  \sum_{i=0}^{rL-1} m_i = 2M$. Hence the two sequences of non-negative integers
  form two partitions, $\lambda$ and $\mu$, respectively, of $2M$. We fill the
  shorter partition with zeros so that both partitions have the same length,
  say $P$.  It is enough to show that $\sum_{i=0}^{i'-1} M_i \leq
  \sum_{i=0}^{i'-1} m_i$ for all $i' = 1, 2, \ldots, P$, i.e., that $\mu$
  majorizes $\lambda$ \cite{Marshall:Olkin:79:1}.

  We show first that $m_0 \geq M_0$. Suppose the contrary, i.e., $m_0 < M_0$. Since
  $m_i \leq m_0$ for all $i \in \{ 0, \ldots, P-1 \}$, we obtain that $m_i <
  M_0$ for all $i \in \{ 0, \ldots, P-1 \}$. But $m_i$, $i \in \{ 0, \ldots,
  P-1 \}$, was obtained by adding over $\Rp$ a certain subset of the set
  $\bigl\{M_{i'} \ | \ i' \in \{0, \ldots P-1 \} \bigr\}$. So there should be
  at least one $m_{i''}$ that has $M_0$ in its composition, and hence $m_{i''}
  \geq M_0$. This is a contradiction, from which we obtain $m_0 \geq M_0$.

  We finish the proof by induction. Namely, we want to show that from
  $\sum_{i=0}^{i'-1} M_i \leq \sum_{i=0}^{i'-1} m_i$ for some $i' \in \{1,
  \ldots P-1 \}$, it follows that $\sum_{i=0}^{i'} M_i \leq \sum_{i=0}^{i'}
  m_i$.  If $M_{i'} \leq m_{i'}$ then this induction step clearly holds. So,
  assume that $M_{i'} > m_{i'}$. Since $m_{P-1} \leq \ldots \leq m_{i'} <
  M_{i'} \leq M_{i'-1} \leq \ldots \leq M_0$, we can deduce that $m_{i'}$, and
  in fact all $m_i$ with $i' \leq i \leq P-1$, cannot contain any $M_i$ with $
  0 \leq i \leq i'$ in its composition. Hence all possible $M_i$, $0 \leq i
  \leq i'$, have occurred in the composition of $m_i$ for $0 \leq i \leq
  i'-1$, which gives $\sum_{i=0}^{i'} m_i \geq \sum_{i=0}^{i'-1} m_i \geq
  \sum_{i=0}^{i'} M_i$.  This proves that $\mu$ majorizes $\lambda$ and we
  obtain 
  \begin{align*}
    \wpsBSC\big( \vomega(D) \big)
      &\geq \wpsBSC\big( \vomega(X) \mod (X^r-1) \big).
  \end{align*}
\end{pr}


Theorem~\ref{th:qc:vs:conv:code:pseudo:weight:1} implies that
low-pseudo-weight vectors in the block code may correspond to higher
pseudo-weight vectors in the convolutional code, but the opposite is not
possible. This suggests that the pseudo-codewords in the block code that
result in decoding failures may not cause such failures in the convolutional
code, thereby resulting in improved performance for the convolutional code at
low-to-moderate signal-to-noise ratios (SNRs). 

A similar bound also holds for the max-fractional weight, as shown in the next
theorem.

\begin{theorem}
  \label{th:qc:vs:conv:code:max:frac:weight:1}
  If $\vomega(D) \in \fc{K}(\matrHconv(D))$, then
  \begin{align*}
    \wmaxfr\big( \vomega(X) \mod (X^r-1) \big)
      &\leq
         \wmaxfr(\vomega(D)).
  \end{align*}
  Therefore,
  \begin{align*}
    \wmaxfrminlr{\matrHQC{r}(X)}
      &\leq
         \wmaxfrmin\big( \matrHconv(D) \big).
  \end{align*}
\end{theorem}

\begin{pr}
 We have $\onenorm{\vomega(D)} = \onenorm{\vomega(X) \mod (X^r-1)}$
 and
 \begin{align*}
   \infnormbig{\vomega(X) \mod (X^r-1)}
     &= \max_{\ell=1}^{L}~
          \max_{i=0}^{r-1}~
            \sum_{i'=0}^{\lfloor (t-1)/r \rfloor}
              \omega_{\ell, i+i'r} 
      \geq
        \max_{\ell=1}^{L}~
          \max_{i=0}^{r-1}~
            \max_{i'=0}^{\lfloor (t-1)/r \rfloor}
              \omega_{\ell, i+i'r}
      \geq
        \infnormbig{\vomega(D)},
 \end{align*}
 which leads to $ \wmaxfr(\vomega(X) \mod (X^r-1)) \leq \wmaxfr(\vomega(D)).
 $ It now follows that $$\wmaxfrmin(\matr{H}_{\mathrm{QC}}^{(\mathrm{r})}) \leq
 \wmaxfrmin(\matrHconv).$$
\end{pr}


In the case of the fractional weight, it is easy to see that for any
$\vomega(D)\in \setV(\fp{P}(\matrHconv(D))) \setminus \{ \vect{0} \}$, we have
$\onenorm{\vomega(D)} = \onenorm{\vomega(X) \mod (X^r-1)}$, and hence
$\wfr(\vomega(X) \mod (X^r-1)) = \wfr(\vomega(D))$.  When comparing the
minimum fractional weight of the convolutional and QC codes, we encounter a
computatinally harder case, as these values must be computed over the set of
nonzero pseudo-codewords that are vertices of the fundamental polytope. This
is not an easy task, because a vertex pseudo-codeword in the convolutional
code might not map into a vertex pseudo-codeword in the QC code.

The theorem below, however, can be established.  For its better understanding,
we recall that $\wfrmin(\matrHQC{r})$ has the following
meaning~\cite{Feldman:03:1, Feldman:Wainwright:Karger:05:1}. Let $\setEQC{r}
\subseteq{\set{I}}(\matrHQC{r})$ be the set of positions where bit flips
occurred when using $\codeCQC{r}$ for data transmission over a BSC with
crossover probability $p$, $0 \leq p < 1/2$. If $|\setEQC{r}| <
\frac{1}{2} \wfrmin(\matrHQC{r})$, then LP decoding is correct.  Similarly,
$\wfrmin(\matrHconv)$ implies the following. If $\setEconv
\subseteq{\set{I}}(\matrHconv)$ is the set of positions where bit flips
occurred when using $\codeCconv$ for data transmission over a BSC, then
$|\setEconv| < \frac{1}{2} \wfrmin(\matrHconv)$ guarantees that LP decoding is
correct.


\begin{theorem}
  \label{th:qc:vs:conv:code:frac:weight:1}
  
  Assume that we are using $\codeCconv$ for data transmission over a BSC with
  cross-over probability $p$, where $0 \leq p < 1/2$, and that bit flips occur
  at positions $\setEconv \subseteq{\set{I}}(\matrHconv)$. If $|\setEconv| <
  \frac{1}{2} \wfrmin(\matrHQC{r})$, then LP decoding is correct. (Note that
  on the right-hand side of the previous inequality we have
  $\wfrmin(\matrHQC{r})$ and not $\wfrmin(\matrHconv)$.)
\end{theorem}


\begin{pr}
  We know that 
  \begin{align}
    \wfrmin\left( \matrHQC{r} \right)
      &\overset{(*)}{\leq}
         \wmaxfrminlr{\matrHQC{r}}
       \overset{(**)}{\leq}
         \wmaxfrminlr{\matrHconv},
           \label{eq:weight:inequalitities:1}
  \end{align}
  where step $(*)$ follows from ~\cite{Feldman:03:1,
    Feldman:Wainwright:Karger:05:1} (see
  also~\cite{Vontobel:Koetter:05:1:subm}) and step $(**)$ follows from
  Theorem~\ref{th:qc:vs:conv:code:max:frac:weight:1}.

  Remember that $\wmaxfrmin(\matrHQC{r})$ has the following
  meaning~\cite{Feldman:03:1, Feldman:Wainwright:Karger:05:1}. Let $\setEconv
  \subseteq{\set{I}}(\matrHconv)$ be the set of positions where the bit flips
  occured when using
  $\codeCconv$ for data transmission over a BSC. If $|\setEconv| < \frac{1}{2}
  \wmaxfrmin(\matrHconv)$, then LP decoding is correct.

  Now, because the theorem statement assumes that $|\setEconv| < \frac{1}{2}
  \wfrmin(\matrHQC{r})$, using~\eqref{eq:weight:inequalitities:1} we have
  $|\setEconv| < \frac{1}{2} \wmaxfrmin(\matrHconv)$ and so, according to the
   meaning of $\wmaxfrmin(\matrHconv)$, LP decoding
  is correct.
\end{pr}
\begin{table}
\caption{The pseudo-weights of the pseudo-codewords in Example~\ref{projecting:conv:pseudo-codewords}.}\label{tablepseudo-weights}
\begin{center}
\begin{tabular}{|c||c|c|c|c|c|}    \hline
                          &           &          &          &           &\\
                          &$\wpsAWGNC$& $\wpsBEC$& $\wpsBSC$& $\wmaxfr$ &$\wfr$\\ \hline\hline&&&&&\\
$ \vomega(D)$             & $8.45$    & $11$     & $6.67$   & $6.5$     & $26$\\\hline
$\vomega(X) \mod (X^r-1)$ &           &          &          &           &\\
{\rm ~for ~all~}$ r\geq 4$& $8.45$    & $11$     & $6.67$   &$6.5$       &$26$ \\\hline&&&&&\\
$\vomega(X) \mod (X^3-1)$ & $7.86$    & $10$     & $6.5$    &$6.5$      &$26$\\\hline &&&&&\\
$\vomega(X) \mod (X^2-1)$ & $6.09$    & $8$      & $5$      & $3.71$       & $26$   \\\hline&&&&&\\
$\vomega(X) \mod (X-1)$   & $3.63$    & $4$      & $3$      & $2.89$       & $26$\\ 
\hline 
\end{tabular}
\end{center}
\end{table}

\begin{remark}
  As discussed at the end of Sec.~6 in~\cite{Vontobel:Koetter:05:1:subm},
  $\wfrminh{H}$ and $\wmaxfrminh{H}$ can give, especially for long codes,
  quite conservative lower bounds on $\wpsBSCminh{H}$. (E.g., the guarantees
  on the error correction capabilities of the LP decoder implied by
  $\wfrminh{H}$ and $\wmaxfrminh{H}$ are not good enough to prove the results
  in~\cite{Feldman:Malkin:Stein:Servedio:Wainwright:04:1}.) However, a
  positive fact about $\wfrminh{H}$ and $\wmaxfrminh{H}$ is that there are
  polynomial-time algorithms that compute these two
  quantities~\cite{Feldman:03:1, Feldman:Wainwright:Karger:05:1}.
\end{remark}

\begin{remark} \label{QC-coverbound} It is not difficult to
  adapt Theorem~\ref{th:qc:vs:conv:code:pseudo:weight:1}and
  Theorem~\ref{th:qc:vs:conv:code:max:frac:weight:1} such that conclusions
  similar to the ones in Theorems~\ref{th:qc:vs:conv:code:pseudo:weight:1}
  and~\ref{th:qc:vs:conv:code:max:frac:weight:1} can be drawn with respect to
  a QC block code with the same structure but a larger circulant size that is
  a multiple of $r$. In fact, most QC block codes with the same structure but
  a larger circulant size, even if not a multiple of $r$, behave according to
  Theorem~\ref{th:qc:vs:conv:code:pseudo:weight:1}.  Using similar arguments
  to the ones in the proofs of these thorems we obtain the following more
  general inequalities that hold for the AWGNC, BEC, BSC max-fractional and
  fractional pseudo-weights.  If $\vomega(D) \in \fc{K}(\matrHconv(D))$,
  then
 $$
  \wps\big(\vomega(X) \mod (X^r-1) \big)\leq \wps\big(\vomega(X) \mod
  (X^{2r}-1) \big)\leq \wps\big(\vomega(X) \mod (X^{4r}-1) \big)\leq \ldots
  \leq \wps\big( \vomega(D) \big), $$
  for all $ r\geq 1$.
In addition, for the AWGNC, BEC, BSC and 
  max-fractional minimum pseudo-weights the following holds for any $m\geq 1$:
  \begin{align*}
    \wps\big(
      \vomega(X) \mod (X^{r}-1) \big)
      &\leq \wps\big( \vomega(X) \mod (X^{mr}-1)\big).
  \end{align*}

\end{remark}

\begin{Example}
  To illustrate how the pseudo-weights of the pseudo-codeword in the
  convolutional code and its projections onto the QC versions satisfy the
  pseudo-weight inequalities in Remark~\ref{QC-coverbound} for all the defined
  pseudo-weights $\wpsAWGNC,\wpsBEC, \wpsBSC,\wmaxfr$ and $\wfr$, we computed
  the pseudo-weights of the pseudo-codewords in
  Example~\ref{projecting:conv:pseudo-codewords}.
  Table~\ref{tablepseudo-weights} contains these results.

\end{Example}



Next we exemplify some of the bounds on the minimum pseudo-weight of codes
proved in this section.  We take a tower of three QC codes together with their
convolutional version and compute their minimum pseudo-weights which,
according to the bounds derived above, form a sequence of increasing numbers,
upper bounded by the minimum pseudo-weight of the convolutional version.
However, due to the large code parameters, we were able to compute only the
minimum pseudo-weight of the smaller code of length 20.  For the other QC
codes we used the methods of~\cite{Koetter:Vontobel:03:1,
  Vontobel:Koetter:05:1:subm} to give lower and upper bounds.


\begin{Example}\label{smallexample2}

  Consider the $(3,4)$-regular QC-LDPC
  code of length $4r$ given by the scalar parity-check matrix, or polynomial
  parity check matrix, respectively,
  \begin{align*}
    \matrHQC{r}
      &= \begin{bmatrix}
           \matr{I}_{0} & \matr{I}_{0} & \matr{I}_{0} & \matr{I}_{0} \\
           \matr{I}_{0} & \matr{I}_{1} & \matr{I}_{2} & \matr{I}_{3} \\
           \matr{I}_{0} & \matr{I}_{4} & \matr{I}_{3} & \matr{I}_{2}
         \end{bmatrix},
   \quad\quad
   \matrHQC{r}(X)
      = \begin{bmatrix}
           1 & 1     & 1     & 1 \\
           1 & X     & X^{2} & X^{3} \\
           1 & X^{4} & X^{3} & X^{2}
    \end{bmatrix}.
  \end{align*}
  \begin{figure}
    \begin{center}
      \epsfig{file=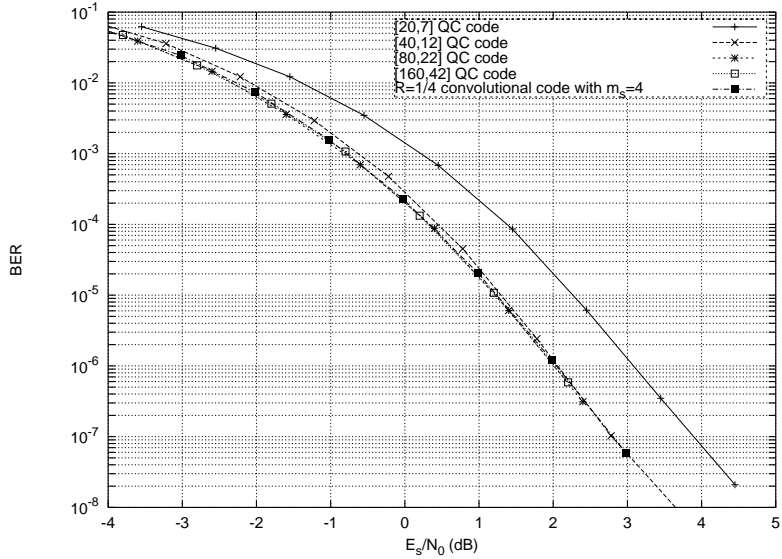, width=0.65\columnwidth}
    \end{center}
    \caption{The performance of a rate $R = 1/4$ $(3,4)$-regular LDPC
      convolutional code and three associated $(3,4)$-regular QC-LDPC block
      codes. Note that the horizontal axis is $\Es/N_0$ and not the more
      common $\Eb/N_0 = (1/R) \cdot \Es/N_0$.}
    \label{fig:compare:small:1}
\end{figure}
  For $r=5$ we obtain a $[20, 7, 6]$ code with rate $R=0.35$. By increasing
  $r$ we obtain other QC codes. By taking powers of $r$ we obtain a tower of
  QC codes whose graphs form a sequence of covers over the Tanner graph of the
  $[20, 7, 6]$ code. For $r\geq 9$, all the codes have minimum distance 10, and
  hence the free distance of the associated rate $R=1/4$ convolutional code is
  $\dfree = 10$, strictly larger than the minimum distance 6 of the $[20,7,
  6]$ code.
  
  For comparison we simulated these codes together with the associated
  convolutional code. The results for an AWGN channel are given in
  Fig.~\ref{fig:compare:small:1}. For the $[20, 7, 6]$ code we ran a vertex
  enumeration program~\cite{Avis:00:1} for polytopes that lists all the minimal
  pseudo-codewords and found that the minimum pseudo-weight of the $[20, 7,
  6]$ code is $\wpsAWGNCmin = 6.00$.  The larger parameters of the other three
  codes allowed us to only lower and upper bound the minimum
  pseudo-weights.\footnote{The lower bounds are obtained by applying the
    techniques that were presented in~\cite{Vontobel:Koetter:04:1}.} For the
  $[40, 12,10]$ QC code, we have $6.05 \leq \wpsAWGNCmin \leq 9.09$ and for
  the $[80, 22,10]$ and the $[160, 42,10]$ codes, we obtained the bounds:
  $6.20 \leq \wpsAWGNCmin \leq 9.09$.\footnote{Using some more sophisticated
    lower bounds from~\cite{Vontobel:Koetter:04:1}, one can actually show that
    $7.19 \leq \wpsAWGNCmin \leq 9.09$ for the length-$40$ code. This implies
    that the lower bounds for the length-80 and length-160 codes can also be
    tightened.} The slight increase in the lower bound from 6, which is the
  minimum weight of the $[20, 7, 6]$ code,  to $6.05$ and $6.20$ shows 
  that the minimum AWGNC pseudo-weight of the $[20, 7, 6]$ code is less than
  that of the $[40, 12,10]$, $[80, 22,10]$, and $[160, 42,10]$ codes. The
  slight increase in the lower bounds from $6.05$ to $6.20$ only suggests, but
  unfortunately gives no further evidence of, the existece of an increasing sequence
  of minimum AWGNC pseudo-weights for these codes, according to the results
  of this section.

  For completeness, we mention that the techniques in~\cite{Feldman:03:1,
    Feldman:Wainwright:Karger:05:1} allow us to efficiently compute the
  minimum max-fractional weight for the above-mentioned codes: we obtain
  $\wmaxfrmin = 4.67$ for the length-$20$ code, $\wmaxfrmin = 5.31$ for the
  length-$40$ code, $\wmaxfrmin = 5.33$ for the length-$80$ code, and
  $\wmaxfrmin = 5.33$ for the length-$160$ code. Applying the results that
  were mentioned in Remark~\ref{remark:relationship:minium:weights:1}, we see
  that these values yield weaker lower bounds on $\wpsAWGNCmin$ than the ones
  given in the previous paragraph.

Using the same vertex enumeration program~\cite{Avis:00:1} we could compute as
well the minimum pseudo-weights of the QC codes of length $4r$ given by the
parity-check matrix $\matrHQC{r}(X)$ in Example~\ref{smallexample2} for $r=1,
2, 3, 4$.  We present the results in Table~\ref{min:pseudoweight:table}.
\begin{table}
\caption{The minimum pseudo-weights of the codes $\codeCQC{r}$, given
  by the parity-check matrix $\matrHQC{r}(X)$, for $r=1,2,3,4.$}\label{min:pseudoweight:table}
\begin{center}
\begin{tabular}{|c||c|c|c|c|} \hline 
                      &              &            &            &             \\ 
$\wpsmin(\codeCQC{r})$&$\wpsAWGNCmin$&$\wpsBECmin$&$\wpsBSCmin$& $\wmaxfrmin$ \\ \hline\hline
$r=1$ & $2 $   & $2 $    & $2$   & $2$ \\\hline
$r=2$ & $2 $   & $2$     & $2$   &$2$  \\\hline 
$r=3$ & $4$    & $4$     & $4$   &$3$  \\\hline 
$r=4$ & $4$    & $4$     & $4$   & $4$  \\\hline
\end{tabular}
\end{center}
\end{table}
\end{Example}

\section{Analysis of Problematic Pseudo-Codewords 
            in Convolutional Codes}
 
\label{sec:analysis:problematic:pseudo:codewords:conv:codes:1}


Studying pseudo-codewords of small pseudo-weight, and, in particular (since
the minimum pseudo-weight is upper bounded by the minimum Hamming weight),
studying pseudo-codewords of pseudo-weight smaller than the minimum Hamming
weight, represents an important problem in the performance analysis of LDPC
codes because it allows us to identify potential failures in MPI decoding.





Upper and lower bounds on the minimum pseudo-weight of a convolutional code
can be obtained by exploiting the ``sliding'' structure of its semi-infinite
parity-check matrix $\matrHconv$ and some of its sub-matrices, which
allows relatively easy computations by taking advantage of the increased
sparseness compared to the corresponding parity-check matrix of an underlying
QC code.  On the one hand, this technique allows us to find certain existing
low-weight pseudo-codewords, and on the other, it illustrates the advantage of using
a convolutional code structure over a block code structure in pseudo-codeword
analysis. In addition, similar to the possible increase in minimum distance
expected when going from a QC code to its unwrapped convolutional version, we
expect an increase in the minimum pseudo-weight of the convolutional code,
leading to better performance compared to the original QC code. Our 
theoretical results and experimental observations point strongly in this
direction. We now explain briefly our technique.
 
Similar to associating with a convolutional code~\cite{Lin:Costello:04:1} an
increasing sequence of {\em column distances} $(\dc_l)_{l\geq 0}$, and a
decreasing sequence of {\em row distances} $(\dr_l)_{l\geq 0},$ having the
property that
\begin{align*}
  \dc_0
    \leq \dc_1
    \leq \cdots
    \leq \dfree 
    \leq \cdots 
    \leq \dr_1
    \leq \dr_{0},
\end{align*}
we define two sequences of pseudo-weights that prove helpful in identifying
the overall minimum pseudo-weight.
 
We recall that an encoder polynomial generator matrix $\matrGconv(D)$ of a
rate $R = K/L$ convolutional code with encoder memory $\me$ has associated
with it a semi-infinite sliding generator matrix $\matrGbarconv$. Let
$\matrGbarconv^{(j,i)}$ denote the $jK \times iL$ submatrix of $\matrGbarconv$
with rows indexed by the first $j$ block rows of $\matrGbarconv$ and columns
indexed by the first $i$ block columns of $\matrGbarconv$, and let
\begin{align*}
  d_{j,i}
    &= \min
         \left\{
           \left.
             \wH\left(
             \vubar \cdot \matrGbarconv^{(j,i)}\right)
           \ \right| \
           \vubar = (\vubar_0, \ldots, \vubar_{j-1}) \in \GF{2}^{jK}, 
           \vubar_0 \neq 0
         \right\}.
\end{align*}
The following two sequences of matrices
\begin{align*}
  \matrGbarconv^{(1,1)}, \
  \matrGbarconv^{(2,2)}, \
  \ldots,
  \matrGbarconv^{(l,l)}, \
  \ldots,
\end{align*}
and
\begin{align*}
  \matrGbarconv^{(1,\me+1)}, \
  \matrGbarconv^{(2,\me+2)}, \
  \ldots,
  \matrGbarconv^{(l,\me+l)}, \
  \ldots,
\end{align*}
give us an increasing sequence of distances $(\dc_l)_{l\geq 0}$, $\dc_l\defeq
d_{l,l}$, commonly called {\em column distances}, associated with the first
sequence, and a decreasing sequence of distances $(\dr_l)_{l\geq 0},$
$\dr_l\defeq d_{l,l+\me}$, commonly called {\em row distances}, associated
with the second sequence.
The column distance $\dc_l$ is a ``truncation'' distance, i.e., it measures
the minimum of the Hamming weights of the vectors of length $l+1$ that
constitute the first $l+1$ components of some codeword with non-zero first
component. The row distance $\dr_l$ is a ``bounded codeword'' distance, i.e.,
it measures the minimum of the Hamming weights of the codewords with non-zero
first component and of length $l+1$ or smaller, or equivalently, of polynomial
degree $l$ or smaller. The column distances and row distances represent
valuable lower bounds and, respectively, upper bounds, on the free distance
that become increasingly tight with increasing $l$, and, in the limit, become
equal to the free distance. If similar sequences could be defined for
pseudo-weights, they would prove helpful in identifying the overall minimum
pseudo-weight.
 
With this in mind, we define corresponding sequences of ``truncated''
pseudo-weights and ``bounded pseudo-codeword'' pseudo-weights.

Let $\matr{H}(D)$ be a polynomial parity-check matrix for a convolutional code
$\codeCconv$ with syndrome former memory $\ms$, and let $\matrHbarconv$ be its
semi-infinite sliding parity-check matrix. Similar to the notation above, let
$\matrHbarconv^{(j,i)}$ be the $jJ \times iL$ submatrix of $\matrHbarconv$
with rows indexed by the first $j$ block rows of $\matrHbarconv$ and columns
indexed by the first $i$ block rows of $\matrHbarconv$. We will consider two
sequences of such sub-matrices:
\begin{align*}
  \matrHbarconv^{(\ms+1,1)}, \
  \matrHbarconv^{(\ms+2,2)}, \
  \ldots,
  \matrHbarconv^{(\ms+l,l)}, \
  \ldots,
\end{align*}
in which $\matrHbarconv^{(\ms+l,l)}$ is the $(\ms+l)J \times lL$
submatrix of the $(\ms+k)J \times kL$ matrix $\matrHbarconv^{(\ms+k,k)}$
formed by its first $lL$ columns, for all $l\leq k$, and
\begin{align*}
  \matrHbarconv^{(1,1)}, \
  \matrHbarconv^{(2,2)}, \
  \ldots,
  \matrHbarconv^{(l,l)}, \
  \ldots,
\end{align*}
in which $\matrHbarconv^{(l,l)}$ is the $lJ \times lL$ submatrix of the $kJ
\times kL$ matrix $ \matrHbarconv^{(k,k)}$ formed by its first $lJ$ rows, for
all $l\leq k$.

In the first sequence, the first matrix that has with certainty a nonzero
nullspace is $\matrHbarconv^{(\ms+\me+1, \me+1)}$, since there is a nonzero
polynomial codeword of degree $\me$ (associated with a scalar codeword of
length $(\me+1)L$). Since $\matrHbarconv^{(\ms+\me+1, \me+1)} 
[\matrGbarconv^{(1,\me+1)}]^\tr=0 $, these matrices act like parity-check
matrices in computing the row distances, with $\matr{H}_{\rm
  conv}^{(\ms+\me+1+l,\me+1+l)}$ giving the $l$-th row distance, for all
$l\geq 0$.  There might be nonzero nullspaces earlier in the sequence, so we
denote by $\matrHbarconv^{(\ms+\mu,\mu)}$ the first matrix with a non-zero
nullspace.  Obviously this would mean that $\matrHconv^{(\mu,\mu)}$ is the
first matrix in the second sequence $\matrHconv^{(l,l)}, l\geq 0$, to have a
nonzero nullspace.  Similarly, $\matrHbarconv^{(l,l)}, l\geq \mu,$ will act
like parity-check matrices in computing the $l$th column distances.  So by
computing the nullspaces of these parity-check matrices we get upper and lower
bounds on $\dfree$ for the convolutional code that are similar to the column
and row distances defined from the generator matrix.
 
We remark also that if we wrap the convolutional code modulo $X^r-1$, with
$r\geq \ms+\mu$, then the matrix $\matrHbarconv^{(\ms+\mu,\mu)}$ is a
submatrix of the parity-check matrix $\matrHQC{r}$ of the QC code that remains
unchanged after the wrapping. Hence a codeword of minimum weight for the
matrix $\matrHbarconv^{(\ms+\mu,\mu)}$ will be, if extended by zeros, a
codeword in the QC code. If this codeword has weight equal to the minimum
distance of the QC code, then the free distance of the convolutional code is
equal to the weight of this codeword. The minimum distance of the QC code
could be smaller, however, and in this case the free distance will only be
upper bounded by the weight of this codeword and lower bounded by the minimum
distance of the QC code.
 
In what follows we will mimic the theory of row distances and column distances
of a convolutional codes to bound the minimum pseudo-weight of the
convolutional code. We obtain upper and lower bounds on the minimum
pseudo-weight of the convolutional code as follows:

\begin{align*}
  \wpsmin\left( \matrHbarconv^{(1,1)} \right)
    &\leq \wpsmin\left(\matrHbarconv^{(2,2)}\right)
     \leq \cdots \\
    &\quad\quad\quad\quad
     \leq \wpsmin\left( \matrHconv \right)
     \leq \cdots 
     \leq \wpsmin \left( \matrHbarconv^{(\ms+2,2)} \right)
     \leq \wpsmin \left( \matrHbarconv^{(\ms+1,1)} \right).
\end{align*}
As above, we will denote by $\matr{H}_{\rm conv}^{(\ms+1,\nu)}$ and
$\matrHbarconv^{(\nu,\nu)}$, respectively, the first matrices in the above
sequences whose fundamental cones contain a non-zero vector.  So by computing
vectors in the fundamental cones of these parity-check matrices we get upper
and lower bounds on $\wpsmin$ for the convolutional
code that are similar to the column and row distances defined from the
generator matrix.

\begin{Example} 
  The pseudo-codeword in Example~\ref{ex:smallexample:1} was obtained by
  attempting to compute small degree non-zero vectors in the fundamental cone
  of $\fc{K}(\matrHbarconv)$ using the above technique. The first nonzero
  ``row pseudo-weight'' is $4$, and the vector $(3D^2+D^3, \ 4D+D^2, \
  3+D+4D^2+D^3, \ 3+4D+D^2)$ is in the fundamental cone of
  $\matrHbarconv^{(8,4)}$. Its AWGN pseudo-weight is $8.45$, which is an upper
  bound on the minimum pseudo-weight of the convolutional code.  The free
  distance of this code is $10$.  The reduced pseudo-codeword $(3X^2+X^3, \
  4X+X^2, \ 3+X+4X^2+X^3, \ 3+4X+X^2)$ modulo $\mod(X^r-1)$, for $r=5, 10, 20,
  40$ has the same weight $8.45$, larger than the minimum distance of the
  $[20, 7, 6]$ code, which makes this pseudo-codeword irrelevant, but smaller
  than the minimum distances of the codes $[40, 12, 10]$, $[80, 22, 10]$,
  $[160, 42, 10]$. The upper bound $\wpsAWGNCmin \leq 9.09$ in Example~\ref{smallexample2}
  therefore becomes $\wpsAWGNCmin
  \leq 8.45$ based on this pseudo-codeword.  
\end{Example} 


This computational method has been applied successfully to larger codes as
well. An example is the rate $R = 2/5$ LDPC-CC with syndrome former memory
$m_\text{s} = 21$ that was simulated in Fig.~\ref{fig:compare:1}. The code was
constructed by unwrapping a $[155, 64]$ $(3,5)$-regular QC-LDPC block code
with minimum Hamming distance $20$. The convolutional code has free distance
$24$,\footnote{The free distance of this convolutional code was obtained by R.
  Johannesson et. al. at Lund University, Dept. of Information Technology,
  using a program called BEAST.} which already suggests a possible performance
improvement compared to the QC code. Following the approach described above,
we constructed a class of pseudo-codewords for which we obtained a minimum
pseudo-weight of $17.85$.  Thus this class of pseudo-codewords contains
vectors of weight less than the free distance, which makes them relevant to
the performance analysis of iterative decoding. Consequently, an upper bound
on the minimum pseudo-weight of the convolutional code is $17.85$, and, from
the way we constructed this class of pseudo-codewords, we believe it is a very
tight bound.  Projecting this pseudo-codeword onto the QC codes obtained by
wrapping the convolutional code gives upper bounds on the minimum
pseudo-weight of these codes as well (in some cases tighter than the ones
obtained using the methods of~\cite{Koetter:Vontobel:03:1,
  Vontobel:Koetter:05:1:subm}). The upper bound
in~\cite{Koetter:Vontobel:03:1, Vontobel:Koetter:05:1:subm} for the minimum
pseudo-weight of the $[155,64]$ code is $16.4$. These upper bounds together
with our simulation curves suggest that the minimum pseudo-weight of the
convolutional code is strictly greater than the minimum pseudo-weight of the
[155,64] QC code. An evaluation of the exact values of the minimum
pseudo-weight in this cases is not possible, however, due to the large
complexity of such a task. Also note that if an upper bound on the minimum
pseudo-weight of the convolutional code smaller than $16.4$ could be found, it
would decrease the upper bound on the minimum pseudo-weight of the $[155,64]$
$(3,5)$-regular QC-LDPC block code as well.

\section{Conclusions}
\label{sec:conclusions:1}


For an LDPC convolutional code derived by unwrapping an LDPC-QC block code, we
have shown that the free pseudo-weight of the convolutional code is at least
as large as the minimum pseudo-weight of the underlying QC code. This result
suggests that the pseudo-weight spectrum of the convolutional code
is "thinner" than that of the block code. This difference in the weight
spectra leads to improved BER performance at low-to-moderate SNRs for the
convolutional code, a conclusion supported by the simulation results presented
in Figs.~\ref{fig:compare:1} and~\ref{fig:compare:small:1}. We also
presented three methods of analysis for problematic pseudo-codewords,
i.e.,~pseudo-codewords with small pseudo-weight. The first method introduces
two sequences of ``truncated'' pseudo-weights and, respectively, ``bounded
pseudo-codeword'' pseudo-weights, which lower and upper bound the minimum
pseudo-weight of the convolutional code, similar to the role that column
distances and row distances play in bounding from below and above the free
distance. The other two methods can be applied to any QC or convolutional code
and consist of projecting codewords with small weight in QC codes or
convolutional codes onto Tanner graph covers of the code.



\end{document}